\documentclass{article}

\usepackage{amstext,amsmath,amssymb}
\usepackage[dvips]{color}
\usepackage{graphicx}
\usepackage{epsfig}
\usepackage{cite}

\begin{document}

\setcounter{secnumdepth}{2}

\title{Efficient model chemistries for peptides. I. Split-valence Gaussian basis sets and the heterolevel approximation in RHF and MP2}

\author{Pablo Echenique$^{1,2}$\footnote{Corresponding author. E-mail
address: {\tt pnique@unizar.es}}\ and
J. L. Alonso$^{1,2}$
\vspace{0.4cm}\\ $^{1}$ {\small Theoretical Physics Department,
University of Zaragoza,}\\
{\small Pedro Cerbuna 12, 50009, Zaragoza, Spain.}\\
$^{2}$ {\small Institute for Biocomputation and Physics of Complex
Systems (BIFI),}\\ {\small Edificio
Cervantes, Corona de Arag\'on 42, 50009, Zaragoza, Spain.}}

\date{\today}

\maketitle

\begin{abstract}
We present an exhaustive study of more than 250 ab initio potential
energy surfaces (PESs) of the model dipeptide HCO-{\small
L}-Ala-NH$_2$. The model chemistries (MCs) used are constructed as
homo- and heterolevels involving possibly different RHF and MP2
calculations for the geometry and the energy. The basis sets used
belong to a sample of 39 selected representants from Pople's
split-valence families, ranging from the small 3-21G to the large
6-311++G(2df,2pd). The reference PES to which the rest are compared is
the MP2/6-311++G(2df,2pd) homolevel, which, as far as we are aware, is
the more accurate PES of a dipeptide in the literature. The aim of the
study presented is twofold: On the one hand, the evaluation of the
influence of polarization and diffuse functions in the basis set,
distinguishing between those placed at 1st-row atoms and those placed
at hydrogens, as well as the effect of different contraction and
valence splitting schemes. On the other hand, the investigation of the
heterolevel assumption, which is defined here to be that which states
that heterolevel MCs are more efficient than homolevel MCs. The
heterolevel approximation is very commonly used in the literature, but
it is seldom checked. As far as we know, the only tests for peptides
or related systems, have been performed using a small number of
conformers, and this is the first time that this potentially very
economical approximation is tested in full PESs. In order to achieve
these goals, all data sets have been compared and analyzed in a way
which captures the nearness concept in the space of MCs.

The most important results of the study are the following: First, that
the convergence in method is not achieved in the \mbox{RHF
$\rightarrow$ MP2} step. Second that the transferability of basis set
accuracy from RHF to MP2 is imperfect. These two conclusions lead us
to discourage the use of RHF MCs for peptides. Regarding the relative
efficiency of the Pople's basis sets, we recommend the inclusion of
polarization functions in 1st-row atoms and we discourage the use of
basis sets containing doubly-split polarization shells and no diffuse
functions. Also, we have found that 6-31G(d) is very efficient for
calculating the geometry, and that both the RHF and MP2 infinite basis
set limits are approximately reached at 6-311++G(2df,2pd). Finally,
related to the heterolevel approximation, we conclude that it is
essentially correct for the description of the conformational behaviour
of HCO-{\small L}-Ala-NH$_2$ both at RHF and MP2. Nevertheless,
we place a cautionary remark on the use of RHF geometries with MP2
single-points: Whereas this practice could be accurate enough for
short peptides, the accumulation of errors may render it unreliable
for longer chains and require the use of MP2 geometries.
\vspace{0.2cm}
\end{abstract}

\section{Introduction}
\label{sec:pess_introduction}

In any bottom-up approach to the still unsolved protein folding
problem \cite{Anf1973SCI,Sko2005PNAS,Dag2003TBS,Hon1999JMB}, the
characterization of the conformational behaviour of short peptides
\cite{Zho2006JCTC,Tor2006MP,DiS2005JCTC,Per2004JCC,Bea1997JACS,Heg2007PRL,Per2003CEJ,Els2000CP}
constitutes an unavoidable first step. If high accuracy of the
treatment is sought, numerically expensive methods must be used to
calculate the physical properties of these protein subunits. This is
why, the most frequently studied peptides are the shortest ones: the
\emph{dipeptides}
\cite{Csa1999PBMB,Kam2007JCTC,Lan2005PSFB,Koo2002JPCA,Hud2001JCC}, in
which a single amino acid residue is capped at both the N- and
C-termini with neutral peptide groups. Among them, the most popular
choice has been the \emph{alanine} dipeptide
\cite{Ros1979JACS,Mez1985JACS,Hea1989IJQC,Per1991JACS,Hea1991JACS,Fre1992JACS,Gou1994JACS,Bea1997JACS,End1997JMST,Rod1998JMST,Els2001CP,Yu2001JMS,Iwa2002JMST,Var2002JPCA,Per2003JCC,Wan2004JCC,Ech2006JCCb,Kam2007JCTC},
which, being the simplest chiral residue, shares many similarities
with most of the rest of dipeptides for the minimum computational
price.

Although classical force fields
\cite{Pon2003APC,Mac1998BOOK,Bro1983JCC,VGu1982MM,Cor1995JACS,Pea1995CPC,Jor1988JACS,Jor1996JACS,Hal1996JCCa}
are the only feasible choice for simulating large molecules at
present, they have been reported to yield inaccurate \emph{potential
energy surfaces} (PESs) for dipeptides
\cite{Kam2007JCTC,Mac2004JCC,Mac2004JACS,Kan2002JMST,Kam2001JPCB,Rod1998JMST}
and short peptides \cite{Wan2006JCTC,Bea1997JACS}. Therefore, it is
not surprising that they are widely recognized as being unable to
capture the fine details needed to correctly describe the intricacies
of the whole protein folding process
\cite{Sno2005ARBBS,Sch2005SCI,Gin2005NAR,Mac2004JCC,Mor2004PNAS,Gom2003BOOK,Kar2002NSB,Bon2001ARBBS}.
On the other hand, albeit prohibitively demanding in terms of
computational resources, ab initio quantum mechanical calculations
\cite{Cra2002BOOK,Jen1998BOOK,Sza1996BOOK} are not only regarded as
the correct physical description that in the long run will be the
preferred choice to directly tackle proteins (given the exponential
growth of computer power), but they are also used in small peptides as
the reference against which less accurate methods must be compared
\cite{Kam2007JCTC,Mau2007JCTC,Arn2006JPCB,Mac2004JCC,Mac2004JACS,Kam2001JPCB,Rod1998JMST,Bea1997JACS}
in order to parameterize improved generations of additive, classical
force fields for polypeptides.

However, despite the sound theoretical basis, in practical quantum
chemistry calculations a plethora of approximations must be typically
made if one wants to obtain the final results in a reasonable human
time. The exact `recipe' that includes all the assumptions and steps
needed to calculate the relevant observables for any molecular system
has been termed \emph{model chemistry} (MC) by John Pople. In his own
words, a MC is an ``approximate but well-defined general and
continuous mathematical procedure of simulation''
\cite{Pop1999RMP}.

The two starting approximations to the exact Schr\"odinger equation
that a MC must contain are, first, the truncation of the $N$-electron
space (in wave\-func\-tion-based methods) or the choice of the
functional (in DFT) and, second, the truncation of the one-electron
space, via the LCAO scheme (in both cases). The extent up to which the
first truncation is carried (or the functional chosen in the case of
DFT) is commonly called the \emph{method} and it is denoted by
acronyms such as RHF, MP2, B3LYP, CCSD(T), FCI, etc., whereas the
second truncation is embodied in the definition of a finite set of
atom-centered Gaussian functions termed \emph{basis set}
\cite{Gar2003BOOK,Jen1998BOOK,Sza1996BOOK,Hel1995BOOK}, which is also
designated by conventional short names, such as 6-31+G(d), TZP or
cc-pVTZ(--f). If we denote the method by a capital $M$ and the basis
set by a $B$, the specification of both is conventionally denoted by
$M/B$ and called a \emph{level of the theory}. Typical examples of
this are RHF/3-21G or MP2/cc-pVDZ
\cite{Cra2002BOOK,Jen1998BOOK,Sza1996BOOK}.

Such levels of the theory are, by themselves, valid MCs,
however, it is very common
\cite{San2005CPL,Per2003JCC,Hob2002JACS,Bea1997JACS} to
use different levels to perform, first, a (possibly constrained)
geometry optimization and, then, a single-point energy calculation on
top of the resulting structures. If we denote by $L_i:=M_i/B_i$ a
given level of the theory, this `mixed' calculation is indicated by
$L_E/\!/L_G$, where the level~$L_E$ at which the single-point energy
calculation is performed is written first
\cite{Pop1999RMP}. Herein, if $L_E \neq L_G$, we shall call
$L_E/\!/L_G$ an \emph{heterolevel} MC; whereas, if $L_E =
L_G$ it will be termed a \emph{homolevel} one, and it will be
typically abbreviated omitting the `double slash' notation.

Apart from the approximations described above, which are the most
commonly used and the only ones that are considered in this work, the
MC concept may include a lot of additional features: protocols for
extrapolating to the infinite-basis set limit
\cite{Jur2006PCCP,Pet2005JCP,Jen2005TCA,Li2001CPL,Nyd1981JCP},
additivity assumptions
\cite{Jur2002CPL,Ign1991JCC,Dew1989JCC,Nob1982CPL}, extrapolations of
the M{\o}ller-Plesset series to infinite order \cite{Pop1983IJQC},
removal of the so-called \emph{basis set superposition error} (BSSE)
\cite{Cre2005JCP,Sen2001IJQC,May1998JCP,Jen1996CPL,May1987TCA,Boy1970MP,Jan1969CPL},
etc. The reason behind most of these techniques is the urging need to
reduce the computational cost of the calculations. For example, in the
case of the heterolevel approximation, this economy principle forces
the level $L_E$ at which the single-point energy calculation is
performed to be more accurate and more numerically demanding than
$L_G$; the reason being simply that, while we must compute the energy
only once at $L_E$, we need to calculate several times the energy and
its gradient with respect to the unconstrained internal nuclear
coordinates at level $L_G$ (the actual number of times depending on
the starting structure, the algorithms used and the size of the
system). Therefore, it would be pointless to use an heterolevel MC
$L_E/\!/L_G$ in which $L_G$ is more expensive than $L_E$, since, at
the end of the geometry optimization, the energy at level $L_E$ is
available.

Now, although general applicability is a requirement that all MCs must
satisfy, general accuracy is not mandatory. Actually, the fact is that
the different procedures that conform a given MC are typically
parameterized and tested in very particular systems, which are often
small molecules. Therefore, the validity of the approximations outside
that native range of problems must be always questioned and checked,
but, while the computational cost of a given MC is easy to evaluate,
its expected accuracy on a particular problem could be difficult to
predict a priori, specially if we are dealing with large molecules in
which interactions in very different energy scales are playing a
role. The description of the conformational behaviour of peptides (or,
more generally, flexible organic species), via their PESs in terms of
the soft internal coordinates, is one of such problems and the one
that is treated in this work.

Our aim here is to provide an exhaustive study of the \emph{Restricted
Hartree Fock} (RHF) and \emph{M{\o}ller-Plesset 2} (MP2) methods,
using the split-valence families of basis sets devised by Pople and
collaborators
\cite{Dit1971JCP,Heh1972JCP,Har1973TCHA,Fri1984JCP,Kri1980JCP,Bin1980JACS,Spi1982JCC,Cla1983JCC}.
To this end, we compare the PES of the model dipeptide HCO-{\small
L}-Ala-NH$_2$ (see fig.~\ref{fig:num_ala_rama}) calculated with a
large number of homo- and heterolevel MCs, and assess their efficiency
by comparison with a reference PES.  Special interest is devoted to
the evaluation of the influence of polarization and diffuse functions
in the basis sets, distinguishing between those placed at 1st-row
atoms and those placed at hydrogens, as well as the effect of
different contraction and valence-splitting schemes.

The second objective of this study, and probably the main one, is the
investigation of the \emph{heterolevel assumption}, which is defined
here to be that which states that \emph{heterolevel MCs are more
efficient than homolevel ones}. The heterolevel assumption is very
commonly assumed in the literature
\cite{Kam2007JCTC,Cur2007JCP,Arn2006JPCB,San2005CPL,Wan2004JCC,Per2003JCC,Iwa2002JMST,Hob2002JACS,Kam2001JPCB},
but it is seldom checked. As far as we know, the only tests for
peptides or related systems, have been performed using a small number
of conformers \cite{Lan2005PSFB,Hud2001JCC,Csa1995JMS,Fre1992JACS,Bea1997JACS},
and this is the first time that this potentially very economical
approximation is tested in full PESs.

In sec.~\ref{subsec:pess_QM_calculations}, the methodological details
regarding the quantum mechanical calculations performed in this work
are provided. In sec.~\ref{subsec:pess_distance}, a brief summary of
the meaning and the properties of the distance introduced in
ref.~\citen{Alo2006JCC} and used for comparing the different MCs is
given for reference. Next, in sec.~\ref{subsec:pess_bs_selection}, we
discuss the rules and criteria that have been used in order to
reasonably sample the enormous space of all Pople's basis sets. In
sec.~\ref{sec:pess_results}, the main results of the investigation are
presented. For convenience, they are organized into four different
subsections: in sec.~\ref{subsec:pess_intraRHF}, a
RHF$/\!/$RHF-intramethod study is performed, whereas the MP2 analogous
is presented in sec.~\ref{subsec:pess_intraMP2}. In
sec.~\ref{subsec:pess_interlude}, a small interlude is dedicated to
reflect on the general ideas and the nearness concept in the space of
MCs that underlie an investigation such as this one, and also to
compare the RHF and MP2 results obtained in the previous two
sections. In sec.~\ref{subsec:pess_inter}, heterolevel MCs in which
the geometry is calculated at RHF and the energy at MP2 are
evaluated. Finally, sec.~\ref{sec:pess_conclusions} is devoted to give
a brief summary of the most important conclusions of the work.

\section{Methods}
\label{sec:pess_methods}

\subsection{Quantum mechanical calculations and internal coordinates}
\label{subsec:pess_QM_calculations}

\begin{figure}[b!]
\begin{center}
\includegraphics[scale=0.10]{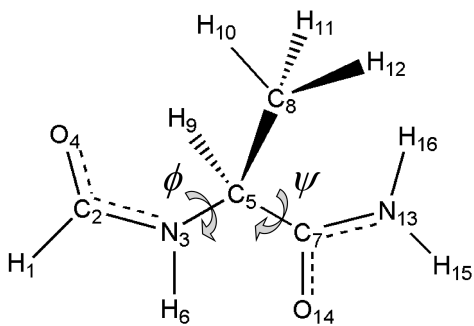}
\end{center}
\caption{\label{fig:num_ala_rama} Atom numeration of the protected
dipeptide HCO-{\small L}-Ala-NH$_2$ according to the SASMIC scheme
introduced in ref.~\citen{Ech2006JCCa}. The soft Ramachandran
angles $\phi$ and~$\psi$ are also indicated.}
\end{figure}

All ab initio quantum mechanical calculations have been performed
using the Gaussian03 program \cite{Gaussian03} under Linux and in
3.20 GHz PIV machines with 2 GB RAM memory. The internal coordinates
used for the Z-matrix of the HCO-{\small L}-Ala-NH$_2$ dipeptide in
the Gaussian03 input files (automatically generated with Perl scripts)
are the \emph{Systematic Approximately Separable Modular Internal
Coordinates} (SASMIC) ones introduced in
ref.~\citen{Ech2006JCCa}. They are presented in
table~\ref{tab:coor_ala_rama} (see also fig.~\ref{fig:num_ala_rama}
for reference). For the geometry optimizations, the SASMIC scheme has
been used too ({\texttt{Opt=Z-matrix}} option) instead of the default
redundant internal coordinates provided by Gaussian03, since we have
seen that, when soft coordinates, such as the Ramachandran angles, are
held fixed and mostly hard coordinates are let vary, the use of the
SASMIC scheme slightly reduces the time to converge with respect to
the redundant internals (for unconstrained optimizations, on the other
hand, the redundant coordinates seem to slightly outperform the SASMIC
ones).

\begin{table}[!t]
\begin{center}
\begin{tabular}{cccc}
Atom name & Bond length & Bond angle & Dihedral angle \\
\hline\\[-8pt]
H$_{1}$ & & & \\
C$_{2}$ & (2,1) & & \\
N$_{3}$ & (3,2) & (3,2,1) & \\
O$_{4}$ & (4,2) & (4,2,1) & (4,2,1,3) \\
C$_{5}$ & (5,3) & (5,3,2) & (5,3,2,1) \\
H$_{6}$ & (6,3) & (6,3,2) & (6,3,2,5) \\
C$_{7}$ & (7,5) & (7,5,3) & $\phi:=${\bf (7,5,3,2)} \\
C$_{8}$ & (8,5) & (8,5,3) & (8,5,3,7) \\
H$_{9}$ & (9,5) & (9,5,3) & (9,5,3,7) \\
H$_{10}$ & (10,8) & (10,8,5) & (10,8,5,3) \\
H$_{11}$ & (11,8) & (11,8,5) & (11,8,5,10) \\
H$_{12}$ & (12,8) & (12,8,5) & (12,8,5,10) \\
N$_{13}$ & (13,7) & (13,7,5) & $\psi:=${\bf (13,7,5,3)} \\
O$_{14}$ & (14,7) & (14,7,5) & (14,7,5,13) \\
H$_{15}$ & (15,13) & (15,13,7) & (15,13,7,5) \\
H$_{16}$ & (16,13) & (16,13,7) & (16,13,7,15)
\end{tabular}
\end{center}
\caption{\label{tab:coor_ala_rama}{\small Internal coordinates in
Z-matrix form of the protected dipeptide HCO-{\small L}-Ala-NH$_2$
according to the SASMIC scheme introduced in
ref.~\citen{Ech2006JCCa}. The numbering of the atoms is that
in fig.~\ref{fig:num_ala_rama}, and the soft Ramachandran angles
$\phi$ and~$\psi$ are indicated.}}
\end{table}

All PESs in this study have been discretized into a regular
12$\times$12 grid in the bidimensional space spanned by the
Ramachandran angles $\phi$ and $\psi$, with both of them ranging from
$-165^{\mathrm{o}}$ to $165^{\mathrm{o}}$ in steps of
$30^{\mathrm{o}}$.

To calculate the geometry at a particular level of the theory, we have
run constrained energy optimizations at each point of the grid,
freezing the two Ramachandran angles $\phi$ and $\psi$ at the
corresponding values, and, in order to save computational resources,
the starting structures were taken, when possible, from PESs
previously optimized at a lower level of the theory. The convergence
criterium for RHF optimizations has been set to {\texttt{Opt=Tight}},
while, in the case of MP2, an intermediate option of
{\texttt{IOp(1/7=100)}} has been used (note that {\texttt{Opt=Tight}}
corresponds to {\texttt{IOp(1/7=10)}}, whereas the default criterium
is {\texttt{IOp(1/7=300)}}). The resulting geometries have been
automatically extracted by Perl scripts and used to construct the
input files for the heterolevel calculations.

The self-consistent Hartree-Fock convergence criterium has been set
in all cases to {\texttt{SCF=(Conver=10)}} (tighter than
{\texttt{SCF=Tight}}) and the MP2 calculations have been performed in
the (default) frozen-core approximation.

At the Hartree-Fock level, 142 PESs have been calculated, taking a
total of $\sim 3.7$ years of computer time, whereas, at MP2, 35 PESs
have been computed and the time invested amounts to $\sim 4.5$ years,
from which, the highest level PES computed in this study, the
MP2/6-311++G(2df,2pd) one depicted in fig~\ref{fig:best_pes_mp2}, has
taken $\sim 3$ years of computer time.  Finally, 88 PESs have been
calculated with MP2$/\!/$RHF-intermethod MCs, taking $\sim 6$
months. In total, 265 PESs of the model dipeptide HCO-{\small
L}-Ala-NH$_2$ have been computed for this study, taking $\sim 8.7$
years of computer time. All these PESs are available as supplementary
material, see the Conclusions for access information.

Finally, let us note that the correcting terms to the PES coming from
mass-metric tensors determinants have been recently shown to be
relevant for the conformational behaviour of peptides
\cite{Ech2006JCCb}. Although, in this study, we have not included
them, the PES calculated here is the greatest part of the effective
free energy \cite{Ech2006JCCb}, so that it may be considered as the
first ingredient for a further refinement of the study in which the
correcting terms are taken into account.

\subsection{Physically meaningful distance}
\label{subsec:pess_distance}

In order to compare the PESs produced by the different (homo- and
heterolevel) MCs, a statistical criterium (distance)
introduced in ref.~\citen{Alo2006JCC} has been used. Let us
recall here that this \emph{distance}, denoted by $d_{12}$, profits
from the complex nature of the system studied to compare any two
different potential energy functions, $V_{1}$ and $V_{2}$. From a
working set of conformations (in this case, the 144 points of each
PES), it statistically measures the typical error that one makes in
the \emph{energy differences} if $V_{2}$ is used instead of the more
accurate $V_{1}$, admitting a linear rescaling and a shift in the
energy reference.

This distance, which has energy units, presents better properties
than other quantities customarily used to perform these comparisons,
such as the energy RMSD, the average energy error, etc., and it may be
related to the Pearson's correlation coefficient $r_{12}$ by

\begin{equation}
\label{eq:pess_d}
d_{12} = \sqrt{2}\,{\sigma}_{2}(1-r_{12}^{2})^{1/2} \  ,
\end{equation}

where $\sigma_2$ is the standard deviation of $V_2$ in the
working set.

Also, due to its physical meaning, it has been argued in
ref.~\citen{Alo2006JCC} that, if the distance between two
different approximations of the energy of the same system is less than
$RT$, one may safely substitute one by the other without altering the
relevant dynamical or thermodynamical behaviour. Consequently, we
shall present the results in units of $RT$ (at
\mbox{$300^{\mathrm{o}}$ K}, so that $RT\simeq 0.6$ kcal/mol).

Finally, if one assumes that the effective energies compared will be
used to construct a polypeptide potential and that it will be designed
as simply the sum of mono-residue ones \cite{Ech2007UNPb}, then, the
number $N_{\mathrm{res}}$ of residues up to which one may go keeping
the distance $d_{12}$ between the two approximations of the the
$N$-residue potential below $RT$ is \cite{Alo2006JCC}

\begin{equation}
\label{eq:chPES_Nres}
N_{\mathrm{res}}=\left ( \frac{RT}{d_{12}} \right )^{2} \ .
\end{equation}

Now, according to the value taken by $N_{\mathrm{res}}$ for a
comparison between a fixed reference PES, denoted by $V_1$, and a
candidate approximation, denoted by $V_2$, we divide all the
efficiency plots in sec~\ref{sec:pess_results} in three regions
depending on the accuracy: the \emph{protein region}, corresponding to
$0 < d_{12} \leq 0.1 RT$, or, equivalently, to $100 \leq
N_{\mathrm{res}} < \infty$; the \emph{peptide region}, corresponding
to $0.1 RT < d_{12} \leq RT$, or $1 \leq N_{\mathrm{res}} < 100$; and,
finally, the \emph{inaccurate region}, where $d_{12} > RT$, and even
for a dipeptide it is not advisable to use $V_2$ as an approximation
to~$V_1$.

\subsection{Basis set selection}
\label{subsec:pess_bs_selection}

In the whole study presented in this work, only Pople's
split-valence basis sets
\cite{Dit1971JCP,Heh1972JCP,Har1973TCHA,Fri1984JCP,Kri1980JCP,Bin1980JACS,Spi1982JCC,Cla1983JCC}
have been investigated. Among the many reasons behind this choice, we
would like to mention the following ones:

\begin{itemize}
\item They are very popular and they are implemented in almost every
 quantum chemistry package, in such a way that they are readily
 available for most researchers and the results here may be easily
 checked or extended.
\item There exist a lot of data calculated using these basis sets in
 the literature, so that the knowledge about their behaviour in
 different problems is constantly growing and may also be enriched by
 the study presented here.
\item Pople's split-valence basis sets incorporate, and hence allow to
 investigate, most of the features and improvements that are commonly
 used in the literature, such as contraction, valence-splitting,
 diffuse functions and polarizations.
\item Unlike some other popular basis sets, such as Dunning's
  correlation-consist\-ent family \cite{Dun1989JCP}, in which the diffuse
 or polarization functions are added in preset groups, Pople's
 basis sets allow for the addition of individual shells rather
 independently, thus permitting a more in depth study. 
\item The number of different basis sets available is very large (see,
 for example the EMSL database at
 {\texttt{http://www.emsl.pnl.gov/forms/basisform. html}}), so that,
 for obvious computational reasons, one cannot explore them all, and
 some choice must be made.
\end{itemize}

Now, even restricting oneself to this particular family of basis
sets, the number of variants that can be formed by independently
adding each type of diffuse or polarization function to each one of
the basic 6-31G and 6-311G sets is huge (to get to the sought point,
there is no need to consider the addition of functions to 3-21G,
4-31G, etc.). Using that the largest set of diffuse and polarization
shells that we may add is the `++G(3df,3pd)' one \cite{Fri1984JCP},
we can express the different basis sets that may be constructed as a
product of all the independent possibilities:

\begin{eqnarray}
\label{eq:chPES_possible_basis_sets}
&& \Big\{ \textrm{6-31G}, \textrm{6-311G} \Big\} \times 
          \Big\{ \,\cdot\, , + \Big\}_{\textrm{1st-row}} \times
          \Big\{ \,\cdot\, , + \Big\}_{\textrm{hydrogen}}
          \times
          \Big\{ \,\cdot\, , \textrm{d} , \textrm{2d} , \textrm{3d}
          \Big\}_{\textrm{1st-row}} \nonumber \\
&& \mbox{} \times
   \Big\{ \,\cdot\, , \textrm{p} , \textrm{2p} , \textrm{3p}
          \Big\}_{\textrm{hydrogen}} \times
   \Big\{ \,\cdot\, , \textrm{f}\, \Big\}_{\textrm{1st-row}}
   \times
   \Big\{ \,\cdot\, , \textrm{d} \Big\}_{\textrm{hydrogen}} \ ,
\end{eqnarray}

where the dot $\,\cdot\,$ indicates here that no function is added
from a particular group.

Therefore, there are $2 \times 2 \times 2 \times 4 \times 4 \times 2
\times 2 = 512$ different Pople's split-valence basis sets just
considering the 6-31G and 6-311G families. This number is
prohibitively large to carry out a full study even at the RHF level,
so that, here, the following strategy has been devised to render the
investigation feasible:

To begin with, we impose several constraints on the basis sets that
will be considered in a first stage:

\begin{table}[!t]
\begin{center}
\begin{tabular}{lll}
\hline\\[-8pt]
 \multicolumn{3}{c}{First-stage, rules-complying basis sets (24)} \\[3pt]
 3-21G        & 6-31G            & 6-311G                  \\
 3-21G(d,p)   & 6-31G(d,p)       & 6-311G(d,p)             \\
 3-21++G      & 6-31G(2d,2p)     & 6-311G(2d,2p)           \\
 3-21++G(d,p) & 6-31G(2df,2pd)   & 6-311G(2df,2pd)         \\
 4-31G        & 6-31++G          & 6-311++G                \\
 4-31G(d,p)   & 6-31++G(d,p)     & 6-311++G(d,p)           \\
 4-31++G      & 6-31++G(2d,2p)   & 6-311++G(2d,2p)         \\
 4-31++G(d,p) & 6-31++G(2df,2pd) & {\bf 6-311++G(2df,2pd)} \\[4pt]
\hline\\[-8pt]
 \multicolumn{3}{c}{First violation of the rules (5)} \\[3pt]
 6-31+G(d,p)            & 6-31++G(d)             & 6-31G(f,d) \\
 6-31$\,\cdot\,$+G(d,p) & 6-31++G($\,\cdot\,$,p) &            \\[4pt]
\hline\\[-8pt]
 \multicolumn{3}{c}{Second violation of the rules (10)} \\[3pt]
 4-31G(d)    & 6-311G(d)    & 6-31G(d)  \\
 4-31+G(d)   & 6-311+G(d)   & 6-31+G(d) \\
 4-31+G(d,p) & 6-311+G(d,p) &           \\
 4-31++G(d)  & 6-311++G(d)  &           \\[4pt]
\hline
\end{tabular}
\end{center}
\caption{\label{tab:basis_sets}{\small Basis sets investigated in this
work. They are organized in three groups: the first one contains the
basis sets that comply with some heuristic restrictions commonly found
in the literature; in the second group, these restrictions are broken
in an exploratory manner; finally, in the third group, 10 new basis
sets are selected according to what has been learned by violating the
rules. The number of basis sets in each group is shown in brackets,
the dot $\,\cdot\,$ is used to indicate that no shell of a particular
type is added to the 1st-row atoms, and the largest
basis set is written in bold face. See also
fig.~\ref{fig:basis_sets}.}}
\end{table}

\begin{enumerate}
\item[(i)] The maximum set of diffuse and polarization shells added is
 `++G(2df, 2pd)', instead of `++G(3df,3pd)'. This is consistent with
 the thumb-rule concept of \emph{balance}~\cite{Jen1998BOOK},
 according to which, the most efficient (\emph{balanced}) basis sets
 are typically those that contain, for each angular momentum $l$, one
 shell more than the ones included for~$l+1$; so that
 6-311++G(3df,3pd), for example, should be regarded as
 \emph{unbalanced}.
\item[(ii)] There must be the same number and type of shells in
 hydrogens as in 1st-row atoms. This has to be interpreted in the
 proper way: for example, if we add two d-type polarization shells to
 1st-row atoms, we must add two p-type ones to hydrogens. They are of
 the same type in the sense that they are one momentum angular unit
 larger than the largest one in the respective valence shell.
\item[(iii)] The higher angular momentum f-type (for 1st-row atoms)
 and d-type shells (for hydrogens), are not included unless the lower
 polarizations are doubly split, i.e., unless we have already included
 the (2d,2p)-shells. This is again consistent with the balance rule
 mentioned in point (i).
\item[(iv)] The investigation of smaller basis sets is restricted to
 the 3-21G and \mbox{4-31G} families, and the largest set of extra shells
 that is added to them is `++G(d,p)'. For consistence, the diffuse and
 polarization functions used for \mbox{3-21G} and 4-31G are the same
 as the ones for 6-31G and \mbox{6-311G}
 \cite{Fri1984JCP,Cla1983JCC,Spi1982JCC,Har1973TCHA}.
\end{enumerate}

\begin{figure}
\begin{center}
\includegraphics[scale=0.25]{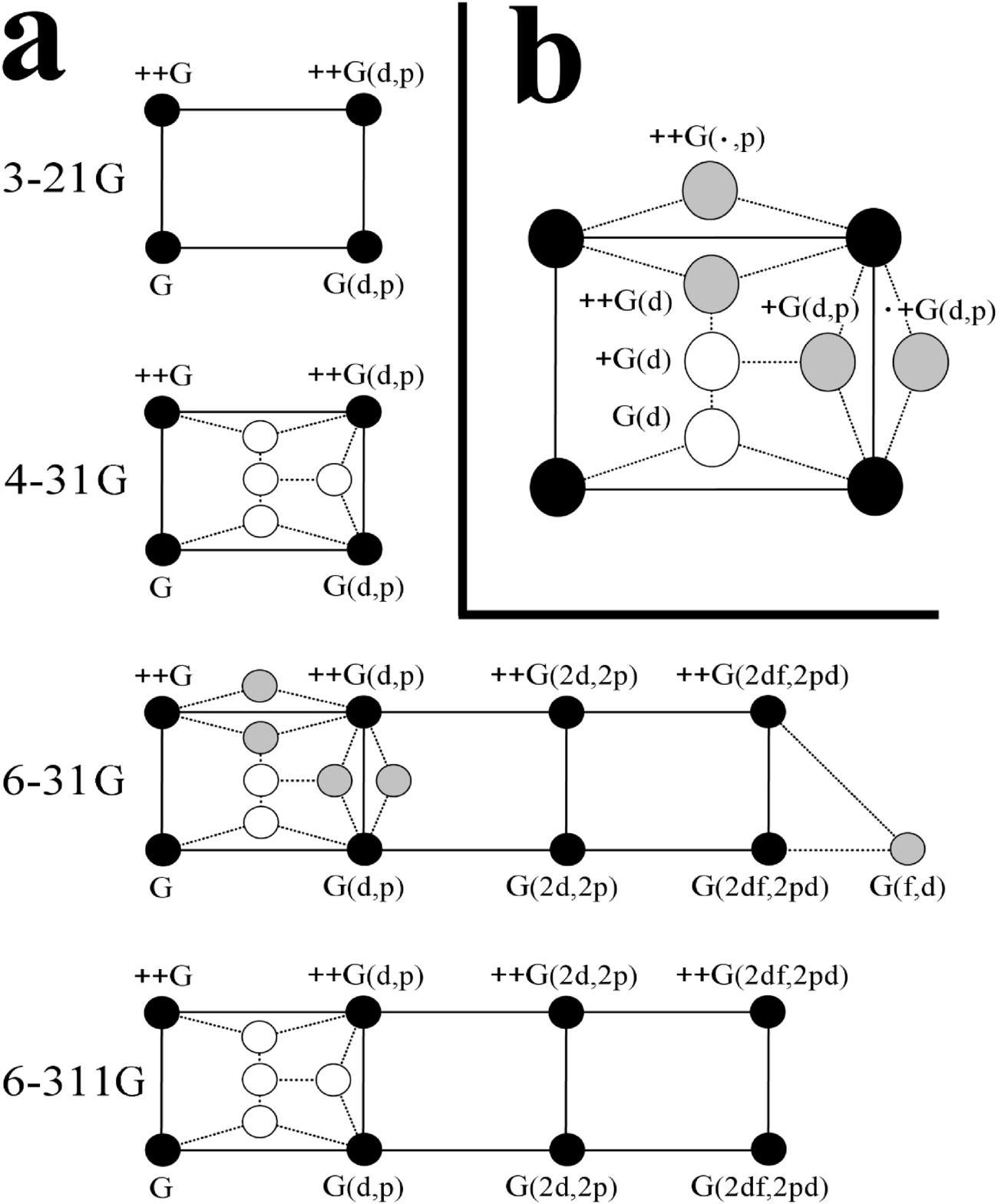}
\caption{\label{fig:basis_sets} Basis sets investigated in this
work. They are organized in three groups: the first one, depicted
as \emph{black circles}, contains the basis sets that comply with some
heuristic restrictions commonly found in the literature; in the second
group, represented as \emph{grey-filled circles}, these restrictions
are broken in an exploratory manner; finally, in the third group,
shown as \emph{white-filled circles}, 10 new basis sets are selected
according to what has been learned by violating the rules. In {\bf
(a)}, a general view of all the 39 basis sets is presented, while in
{\bf (b)}, the left-most region of the 6-31G family has been enlarged
so that the basis sets belonging to the second and third groups could
be more easily appreciated. The dot $\,\cdot\,$ is used to indicate
that no shell of a particular type is added to the 1st-row atoms, and
the geometric arrangement of the basis sets has no deep meaning
whatsoever, it is only intended to provide visual comfort. See also
table~\ref{tab:basis_sets}.}
\end{center}
\end{figure}

These \emph{rules}, most of which are typically obeyed (often tacitly)
in the literature
\cite{Ech2006JCCb,Lan2005PSFB,Wan2004JCC,Per2004JCC,Per2003CEJ,Per2003JCC,Iwa2002JMST,Yu2001JMS,Hud2001JCC,End1997JMST,Gou1994JACS,Fre1992JACS,Ign1991JCC},
produce the list of 24 basis sets labeled as `First-stage,
rules-complying basis sets' in table~\ref{tab:basis_sets} and depicted
as black circles in fig.~\ref{fig:basis_sets}.

Even if their exhaustive study is already a demanding computational
task and the space of all Pople's split-valence basis sets may be
thought to be reasonably sampled by this `first-stage' group, we
wanted to check the validity of some of the rules, since, in the same
spirit of the arguments given in the introduction, what is good for a
particular system or a particular purpose is not necessarily good for
a different one, which may be far away from the native playground
where the methods have been traditionally tested and parameterized.
Therefore, to this end, we have chosen the medium-sized and reasonably
RHF-efficient 6-31++G(d,p) basis set (see
sec.~\ref{subsec:pess_intraRHF}), and we have modified it in order to
break restrictions~(ii) and~(iii). On the one hand, as representants
of breaking rule~(ii), we have selected the basis sets
\mbox{6-31+G(d,p)}, 6-31++G(d), 6-31$\,\cdot\,$+G(d,p) and
6-31++G($\,\cdot\,$,p), where, in the first two cases, a diffuse and a
polarization shell has been respectively removed from the hydrogens,
while, in the last two ones, the removal has been carried out on the
1st-row atoms. This second modification is so unusual (in fact, we
have not found any work where it is performed) that there is no
notation for it in the literature; herein, a dot $\,\cdot\,$ is used
in the place where the unexisting 1st-row-atom shell would appear. On
the other hand, as a representant of breaking rule~(iii), we have
selected 6-31G(f,d). This new group of 5 basis sets is labeled as
`First violation of the rules' in table~\ref{tab:basis_sets} and
depicted as grey-filled circles in fig.~\ref{fig:basis_sets}. We have
decided to violate neither rule~(i), mainly for computational reasons,
nor rule (iv), due to the fact that the study of the smaller basis
sets is intended to be only exploratory (and, in any case, the 3-21G
and 4-31G families have proved to be rather inefficient for this
problem; see sec.~\ref{sec:pess_results}).

The conclusions extracted from the study of the `first violation of
the rules' group within RHF are discussed later, however, it suffices
to say for the moment that we learn from them that breaking rule~(iii)
is not advantageous, and that one may benefit from breaking rule~(ii)
only if the functions are removed from the hydrogens. Therefore, in
the final step of the selection of the basis sets that shall be
investigated, we include a new group of 10 basis sets which come from
removing hydrogen-atom diffuse and/or polarization shells from some of
the most efficient ones in the other two groups. This new block is
labeled as `Second violation of the rules' in
table~\ref{tab:basis_sets} and depicted as white-filled circles in
fig.~\ref{fig:basis_sets}.

The basis sets used in the RHF part of the study are those in
table~\ref{tab:basis_sets}, whereas, in the MP2 part, we have
considered the smaller subgroup that may be found in
table~\ref{tab:basis_sets_MP2} (see also
fig.~\ref{fig:basis_sets}). All of them have been taken from the
Gaussian03 internally stored library except for
\mbox{6-31$\,\cdot\,$+G(d,p)}, \mbox{6-31++G($\,\cdot\,$,p)} and all
the basis sets generated from the 3-21G and 4-31G ones by adding extra
functions. The first two have no accepted notation and cannot be
specified in the program, while the ones derived from 3-21G and 4-31G
have been constructed, for consistence, using the diffuse and/or
polarization shells of the 6-31G and 6-311G families. For these
exceptions, the data has been taken from the EMSL repository at
{\texttt{http://www.emsl.pnl.gov/forms/basisform.html}}\footnote{
Basis sets were obtained from the Extensible Computational Chemistry
Environment Basis Set Database at
\texttt{http://www.emsl.pnl.gov/forms/basisform.html}, Version
02/25/04, as developed and distributed by the Molecular Science
Computing Facility, Environmental and Molecular Sciences Laboratory
which is part of the Pacific Northwest Laboratory, P.O. Box 999,
Richland, Washington 99352, USA, and funded by the U.S. Department of
Energy. The Pacific Northwest Laboratory is a multi-program laboratory
operated by Battelle Memorial Institute for the U.S. Department of
Energy under contract DE-AC06-76RLO 1830. Contact Karen Schuchardt for
further information.} and the basis sets have been read using the
{\texttt{Gen}} keyword. In all cases, spherical Gaussian-type orbitals
(GTOs) have been preferred, thus having 5 d-type and 7 f-type
functions per shell.

\section{Results}
\label{sec:pess_results}

\subsection{RHF$/\!/$RHF-intramethod model chemistries}
\label{subsec:pess_intraRHF}

The study in this work begins by performing an exhaustive comparison
of all the homolevel MCs and most of the heterolevel ones that can be
constructed using the 39 different basis sets described above and
within the RHF method. The original aim was to identify, separately,
the most efficient basis sets for doing geometry optimizations and
those that perform best for single-point energy calculations, in order
to extract the information needed to carry out, in successive stages,
a (necessarily) more restrictive study of MP2$/\!/$MP2 and
MP2$/\!/$RHF MCs. However, due to the considerations made in
sec.~\ref{subsec:pess_interlude}, all mentions to the accuracy of any
given MC in this section must be understood as relative to the
RHF$/\!/$RHF reference, and not to the (surely better) MP2$/\!/$MP2
one or to the exact result. In this spirit, this part of the study
should be regarded as an evaluation of the most efficient MCs for
approximating \emph{the infinite basis set Hartree-Fock limit} (for
which the best RHF$/\!/$RHF homolevel here is probably a good
approximation), and also as a way of introducing the relevant concepts
and the systematic approach that shall be used in the rest of the
computationally more useful sections.

Having this in mind, the \emph{efficiency} of a particular MC is laxly
defined as a balance between accuracy (in terms of the distance
introduced in sec.~\ref{subsec:pess_distance}) and computational cost
(in terms of time). It is graphically extracted from the
\emph{efficiency plots}, where the distance $d_{12}$ between any given
MC and a reference one is shown in units of $RT$ in the $x$-axis,
while, in the $y$-axis, one can find in logarithmic scale the average
computational time taken for each MC, per point of the 12$\times$12
grid defined in the Ramachandran space of the model dipeptide
HCO-{\small L}-Ala-NH$_2$ (see the following pages for several
examples). As a general thumb-rule, \emph{we shall consider a MC to be
more efficient for approximating the reference when it is placed
closer to the origin of coordinates in the efficiency plot}. This
approach is intentionally non-rigorous due to the fact that many
factors exist that influence the computer time but may vary from one
practical calculation to another; such as the algorithms used, the
actual details of the computers (frequency of the processor, size of
the RAM and cache memories, system bus and disk access velocity,
operating system, mathematical libraries, etc.), the starting guesses
for the SCF orbitals or the starting structures in geometry
optimizations. Taking this into account, the only conclusions that
shall be drawn in this work about the relative efficiency of the MCs
studied are those deduced from strong signals in the plots and,
therefore, those that can be extrapolated to future calculations; in
other words, \emph{the small details shall be typically neglected}.

\begin{figure}
\begin{center}
\includegraphics[scale=0.40]{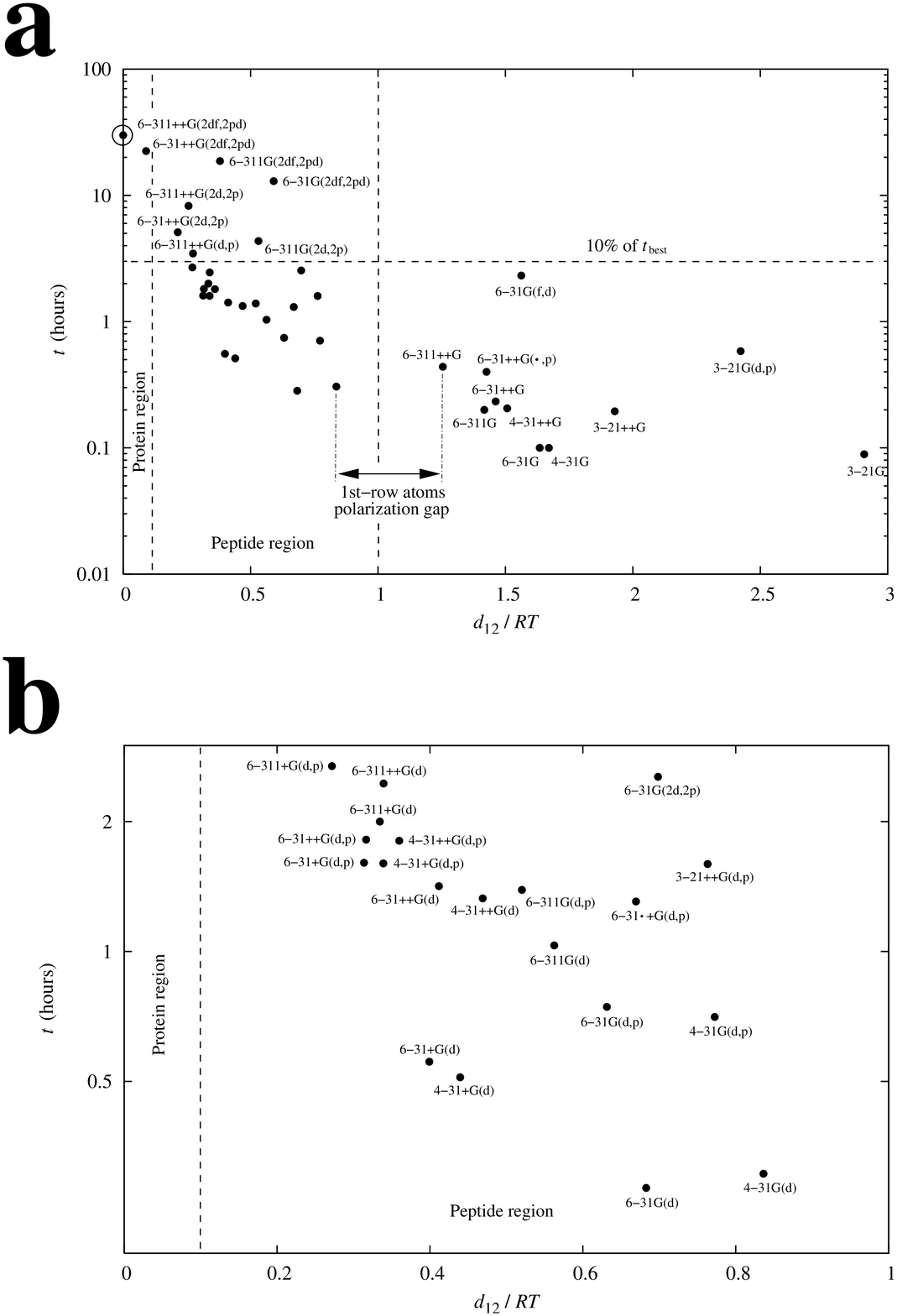}
\caption{\label{fig:distance_homo_RHF} Efficiency plots of the
\emph{RHF-homolevel} MCs corresponding to the basis sets in
table~\ref{tab:basis_sets}. In the $x$-axis, we show the distance
$d_{12}$, in units of $RT$ at \mbox{$300^{\mathrm{o}}$ K}, between any
given MC and the reference one (indicated by an encircled point),
while, in the $y$-axis, we present in logarithmic scale the average
computational time taken for each MC, per point of the 12$\times$12
grid defined in the Ramachandran space of the model dipeptide
HCO-{\small L}-Ala-NH$_2$. {\bf (a)}~General view containing all basis
sets. {\bf (b)} Detailed zoom of the most efficient region of the plot
($d_{12} < RT$ and \mbox{$t < 10\%$ of $t_{\mathrm{best}}$}).}
\end{center}
\end{figure}

\begin{figure}
\begin{center}
\includegraphics[scale=0.39]{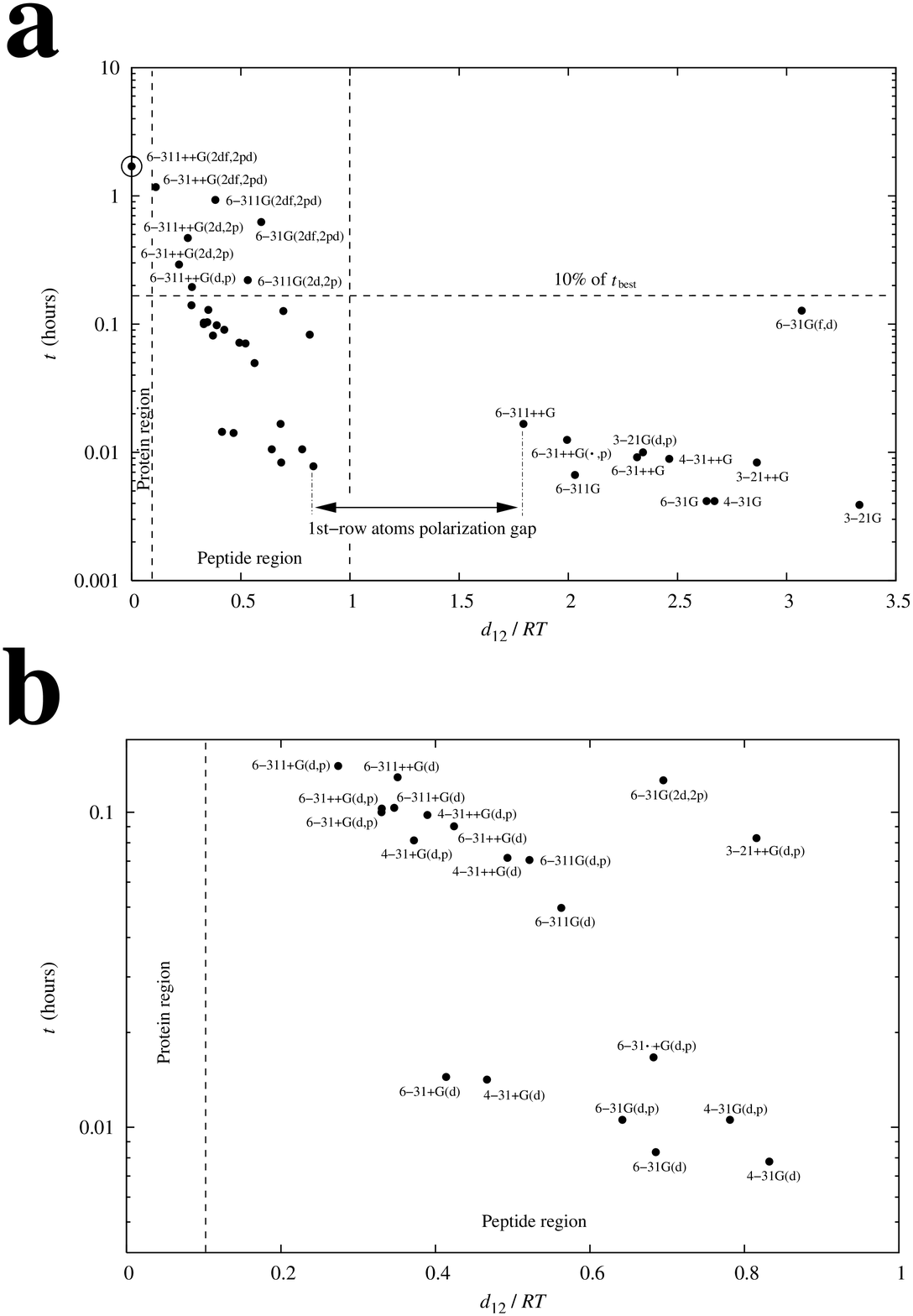}
\caption{\label{fig:distance_hetero_sp_RHF} Efficiency plots of the
\emph{RHF-heterolevel} MCs $L_E^{i} /\!/ L_G^{\mathrm{best}}$ obtained
computing the geometry at the best level of the theory,
$L^{\mathrm{best}}:=$6-311++G(2df,2pd), and then performing a
single-point calculation with all the basis sets in
table~\ref{tab:basis_sets} but the largest one. In the $x$-axis, we
show the distance $d_{12}$, in units of $RT$ at
\mbox{$300^{\mathrm{o}}$ K}, between any given MC and the reference
one (the \emph{homolevel} \mbox{6-311++G(2df,2pd)}, indicated by an
encircled point), while, in the $y$-axis, we present in logarithmic
scale the average computational time taken for the corresponding
single-point, per point of the 12$\times$12 grid defined in the
Ramachandran space of the model dipeptide HCO-{\small
L}-Ala-NH$_2$. {\bf (a)}~General view containing all basis sets. {\bf
(b)} Detailed zoom of the most efficient region of the plot ($d_{12} <
RT$ and \mbox{$t < 10\%$ of $t_{\mathrm{best}}$}).}
\end{center}
\end{figure}

The efficiency plots that we will discuss in this section are the ones
used to compare \emph{RHF$/\!/$RHF-intramethod} homo- and heterolevel
MCs with the RHF \emph{reference}, defined as the homolevel MC with
the largest basis set, i.e., RHF/ 6-311++G(2df,2pd) (since, in
this section, there is no possible ambiguity, the levels shall be
denoted in what follows omitting the `RHF' keyword and specifying only
the basis set). The plots corresponding to this first intramethod part
comprise figures from~\ref{fig:distance_homo_RHF}
to~\ref{fig:distance_selected_and_everything_RHF}.

In fig.~\ref{fig:distance_homo_RHF}, the \emph{homolevel} MCs
corresponding to all the basis sets in table~\ref{tab:basis_sets} are
compared to the reference one. In fig.~\ref{fig:distance_homo_RHF}a, a
general picture is presented, whereas, in
fig.~\ref{fig:distance_homo_RHF}b, a detailed zoom of the most
efficient region of the plot is shown. It takes an average of~$\sim
30$ hours per grid \label{foot:howtimegeo}point to calculate the PES
of the model dipeptide HCO-{\small L}-Ala-NH$_2$ at the reference
homolevel 6-311++G(2df,2pd) (the time per point for homolevels is
calculated assuming that all geometry optimizations take 20 iterations
to converge, this is done in order to avoid the ambiguity due to the
choice of the starting structures and it allows to place all MCs on a
more equivalent footing); this time is denoted by~$t_{\mathrm{best}}$
and the most efficient region is defined as that in which $d_{12} <
RT$ and \mbox{$t < 10\%$ of $t_{\mathrm{best}}$}. Additionally, we
indicate in the plots the \emph{peptide region} ($0.1 RT < d_{12} \leq
RT$), containing the MCs that may correctly approximate the reference
one for chains of 1--100 residues, and the \emph{protein region} ($0 <
d_{12} \leq 0.1 RT$), including the MCs that are accurate for
polypeptides over 100 residues (see sec.~\ref{sec:pess_methods}).

From these two plots, several conclusions may be drawn:

\begin{itemize}
\item Regarding the check of rules (ii) and (iii) via the basis sets
 in the second group in table~\ref{tab:basis_sets}, we see that
 6-31+G(d,p) is more efficient than 6-31++G(d,p) (it is cheaper and,
 despite being smaller, more accurate!\footnote{ Note that the
 Hartree-Fock method has a variational origin, in such a way that, if
 we were interested in the absolute value of the energy, and not in
 the energy differences, an enlargement of the basis set would always
 improve the results.}), that 6-31++G(d) is as efficient as the most
 efficient basis sets of the rules-complying group (being outperformed
 only by some of the ones in the third group in
 table~\ref{tab:basis_sets}), that 6-31$\,\cdot\,$+G(d,p) has drifted
 a little towards the inefficiency region and that that
 6-31++G($\,\cdot\,$,p) is well deep in it. This suggests that
 \emph{it may be profitable to break rule}~(ii) \emph{but only in the
 direction of removing shells from the hydrogens, and not from the
 1st-row atoms}; in agreement with the common practice in the
 literature \cite{Ech2006JCCa,Tor2006MP,Wan2004JCC,Per2003JCC,Koo2002JPCA,Iwa2002JMST,Els2001CP,Top2001JACS,Hea1991JACS,Hea1989IJQC}
 based on the intuition that `hydrogens are typically more passive
 atoms sitting at the end of bonds' \cite{Jen1998BOOK}. On the
 other hand, 6-31G(f,d) turns out to be very inefficient, being about
 as accurate as the simple 6-31G basis set but far more
 expensive. This confirms that \emph{it is a good idea to follow
 restriction}~(iii), which is consistent with the already mentioned
 thumb-rule of basis set `balance' \cite{Jen1998BOOK}.
\item \emph{The whole} 3-21G \emph{family of basis sets is very
 inefficient}. Only 3-21++G(d,p) lies in the accurate region and,
 anyway, it is less efficient than most of the other basis sets there.
\item Contrarily, \emph{the} 4-31G \emph{family results are quite
 parallel to and only slightly worse than those of the} 6-31G
 \emph{family}, suggesting that, to account for conformational energy
 differences within the RHF method, the contraction of valence
 orbitals is more important than the contraction of core ones if
 homolevel MCs are used.
\item In fig.~\ref{fig:distance_homo_RHF}a, we can notice the
 existence of a \emph{gap} in the values of the distance $d_{12}$,
 which lies around $d_{12} = RT$ and separates the MCs in two
 groups. Notably, all the basis sets in the most accurate group share
 a common characteristic: \emph{they contain 1st-row atoms
 polarization functions}, whereas those in the inaccurate group do
 not, with the only exceptions of 3-21G(d,p) and 6-31G(f,d), whose bad
 quality has been explained in the previous points for other reasons.
\item \emph{All the basis sets with extra polarizations}, (2d,2p)
 \emph{or} (2df,2pd), \emph{and no diffuse functions are less
 efficient than their diffuse functions-containing counterparts.}
\item Out of some of the specially inefficient cases discussed in the
 preceding points, \emph{the addition of diffuse functions to
 singly-polarized} ((d) or (d,p)) \emph{basis sets always increases
 the accuracy}.
\item \emph{The only basis set whose homolevel MC lies in the protein
 region is the expensive} 6-31++G(2df,2pd).

\begin{figure}
\begin{center}
\includegraphics[scale=0.38]{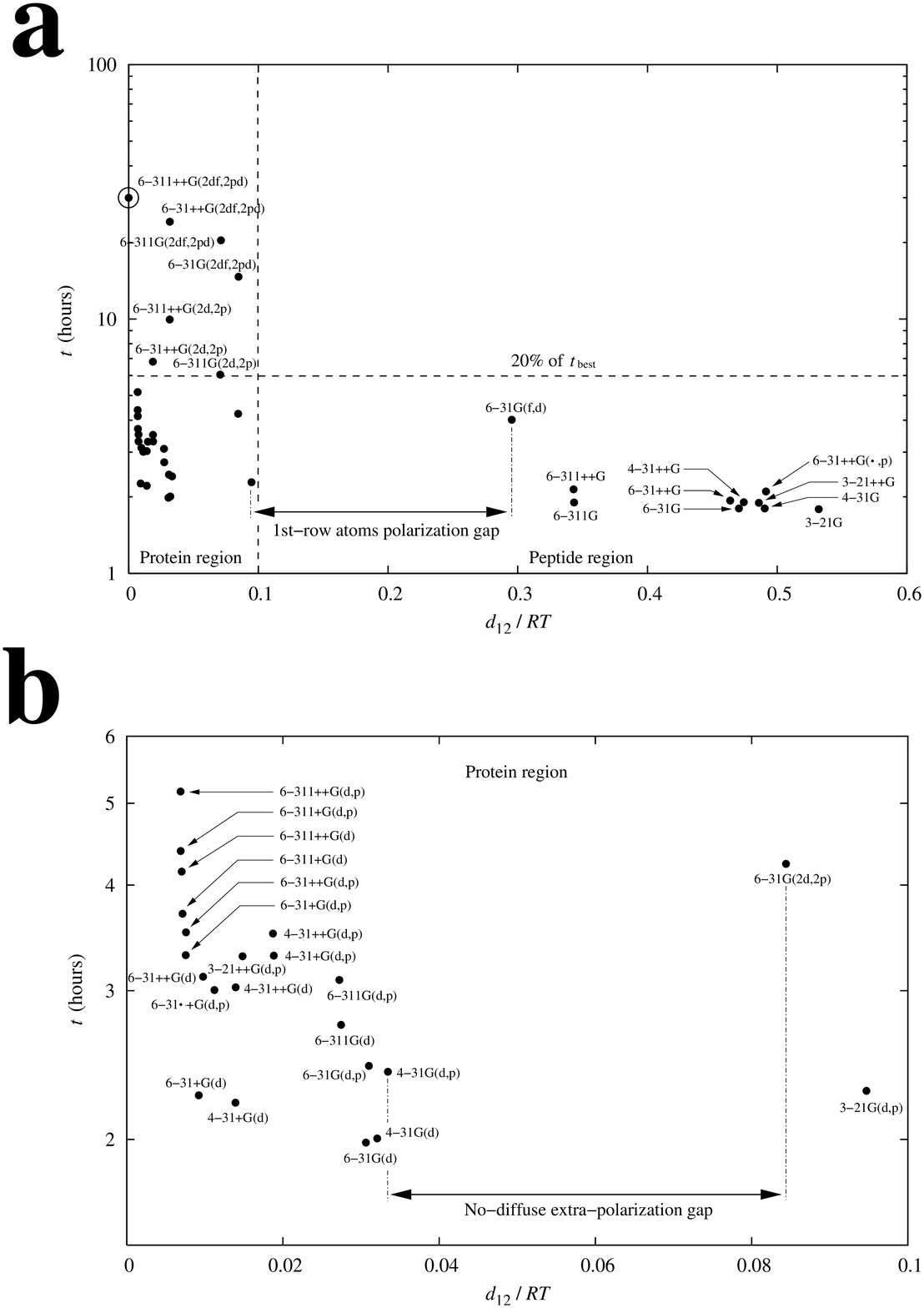}
\caption{\label{fig:distance_hetero_geometry_RHF} Efficiency plots of
the \emph{RHF-heterolevel} MCs $L_E^{\mathrm{best}} /\!/ L_G^{i}$
obtained computing the geometries with all the basis sets in
table~\ref{tab:basis_sets} but the largest one and then performing a
single-point energy calculation at the best level of the theory,
$L^{\mathrm{best}}:=$6-311++G(2df,2pd), on top of each one of them. In
the $x$-axis, we show the distance $d_{12}$, in units of $RT$ at
\mbox{$300^{\mathrm{o}}$ K}, between any given MC and the reference
one (the \emph{homolevel} 6-311++G(2df,2pd), indicated by an encircled
point), while, in the $y$-axis, we present in logarithmic scale the
average computational time taken for each MC, per point of the
12$\times$12 grid defined in the Ramachandran space of the model
dipeptide HCO-{\small L}-Ala-NH$_2$. {\bf (a)}~General view containing
all basis sets. {\bf (b)}~Detailed zoom of the most efficient region
of the plot ($d_{12} < RT$ and \mbox{$t < 20\%$ of
$t_{\mathrm{best}}$}).}
\end{center}
\end{figure}

\item If we look at the most efficient basis sets (those that lie at
 the lower-left envelope of the `cloud' of points), we can see that
 \emph{no accumulation point is reached}, i.e., that, although the
 distance between 6-311++G(2df,2pd) and 6-31++G(2df,2pd) is small
 enough to suggest that we are close to the Hartree-Fock limit for
 this particular problem, if the basis set is intelligently enlarged,
 we obtain increasingly better MCs.
\item For less than 10\% the cost of the reference calculation, some
 particularly efficient basis sets for RHF-homolevel MCs
 that can be used without altering the relevant conformational
 behaviour of short peptides (i.e., whose distance $d_{12}$ with
 6-311++G(2df,2pd) is less than $RT$) are 6-31+G(d,p), 6-31+G(d),
 \mbox{4-31+G(d)} and 6-31G(d).
\end{itemize}

Next, in fig.~\ref{fig:distance_hetero_geometry_RHF}, the reference
homolevel 6-311++G(2df,2pd) is compared to the
\emph{RHF$/\!/$RHF-intramethod-heterolevel} MCs $L_E^{\mathrm{best}}
/\!/ L_G^{i}$ obtained computing the geometries with the~38 remaining
basis sets in table~\ref{tab:basis_sets} and then performing a
single-point energy calculation at the best level of the theory,
$L^{\mathrm{best}}:=$ 6-311++G(2df,2pd), on top of each one of the
structures. The aim of this comparison is twofold: on the one hand, we
want to measure the relative efficiency of the different basis sets
for calculating the \emph{geometry} (not the energy), on the other
hand, we want to find out whether or not the \emph{heterolevel
approximation} described in the introduction is an efficient one
within RHF.

Like in the previous case, in
fig.~\ref{fig:distance_hetero_geometry_RHF}a, a general picture is
presented, whereas, in fig.~\ref{fig:distance_hetero_geometry_RHF}b, a
detailed zoom of the most efficient region of the plot is shown. The
average time per point $t$ of the heterolevel MCs has been calculated
adding the average cost of performing a single-point at
$L^{\mathrm{best}}:=$6-311++G(2df,2pd) ($\sim 1.7$ hours) to the
average time per point needed to calculate the geometry at each one of
the levels $L_{G}^i$ (see page~\pageref{foot:howtimegeo}). This $\sim
1.7$ hours `offset' in all the times, has rendered advisable to set
the limit used to define the efficient region in this case to the 20\%
of $t_{\mathrm{best}}$ (instead of the former 10\%), so that most of
the relevant basis sets are included in the second plot in
fig.~\ref{fig:distance_hetero_geometry_RHF}b.

In this second part of the present section, several interesting
conclusions may be extracted from the plots:

\begin{itemize}
\item Related to the test of rules (ii) and (iii), similar remarks to
 the ones before can be made, the only difference being that, in this
 case, for computing the geometry, 6-31$\,\cdot\,$+G(d,p) is not as
 bad as for the homolevel calculation. The signal, however, is rather
 weak and \emph{the main conclusions stated in the first point above
 should not be modified}.
\item Regarding the 3-21G family of basis sets, we see here that,
 differently from what happened for the homolevels, \emph{they are not
 so bad to account for the geometry}, and, in the case of 3-21G,
 3-21++G and 3-21++G(d,p), their efficiency is close to that of the
 corresponding 4-31G and 6-31G counterparts.
\item \emph{Moreover, the} 4-31G \emph{basis sets performance is still
 quite close to that of the} \mbox{6-31G} \emph{family}. This point,
 together with the previous one, and differently from what happened in
 the case of homolevel MCs, suggests that, to account
 for the equilibrium geometry within RHF, the contraction scheme is
 only mildly important both for valence and core orbitals.
\item In fig.~\ref{fig:distance_hetero_geometry_RHF}a, we see again,
 a \emph{gap} in the values of the distance $d_{12}$ separating the
 MCs with the geometry calculated using basis sets that contain
 1st-row atoms polarization functions from those that do
 not. The only differences are that, this time, the gap is even more
 evident, it lies around $d_{12} = 0.2 RT$, and
 3-21G(d,p) is placed below it.
\item The signal noticed in the homolevel case regarding the relative
 inefficiency of the the basis sets with \emph{extra polarizations},
 (2d,2p) \emph{or} (2df,2pd), \emph{and no diffuse functions} has
 become stronger here and a second \emph{gap} can be seen separating
 them from their diffuse functions-containing counterparts and also
 from the basis sets with only one polarization shell. This gap
 separates, for example, the MCs whose geometries have been calculated
 with 6-31G(2df,2pd) and 6-31G(d,p), in such a way that the smaller
 one is not only more efficient, \emph{but also more accurate}. This
 clearly illustrates one of the points raised in this study, namely,
 that MCs parameterized and tested in concrete systems may behave in
 an unexpected way when used in a different problem, and that the
 investigation of the quality of the most popular MCs, as well as the
 design of new ones, for the study of the conformational preferences
 of peptides, is a necessary (albeit enormous) task.
\item Also for geometry optimizations, \emph{the addition of diffuse
 functions to singly-polarized} ((d) or (d,p)) \emph{basis sets
 increases the accuracy}.
\item Contrarily to the situation for homolevels, where the only basis
 set that lied in the protein region was the 6-31++G(2df,2pd) one and
 some MCs presented distances of near $3 RT$ with the reference one,
 here, all MCs lie well below $d_{12} = RT$, and those for which the
 geometry has been computed with a basis set that contains 1st-row
 atoms polarization functions (except for \mbox{6-31G(f,d)}) are
 \emph{all in the protein region}, so that, under the assumptions in
 sec.~\ref{subsec:pess_distance}, they can correctly approximate the
 reference~MC for chains of more than 100 residues. Remarkably, some
 of this heterolevel~MCs, such as
 \mbox{6-311++G(2df,2pd)$/\!/$6-31+G(d)} for example, are physically
 equivalent to the reference homolevel up to peptides of \emph{ten
 thousand residues} at less of~10\% the computational cost. Indeed,
 all these results \emph{confirm the heterolevel assumption},
 discussed in the introduction and so commonly used in the literature
 \cite{Kam2007JCTC,Cur2007JCP,Arn2006JPCB,San2005CPL,Wan2004JCC,Per2003JCC,Iwa2002JMST,Hob2002JACS,Kam2001JPCB}, for
 RHF$/\!/$RHF-intramethod MCs.
\item Differently from the homolevel case, \emph{an accumulation point
 is reached} here in the basis sets, since, in
 fig.~\ref{fig:distance_hetero_geometry_RHF}b, we can see that there
 is no noticeable increase in the accuracy, say, beyond 6-311+G(d).
\item Finally, let us mention 6-311+G(d), 6-31+G(d) and
 \mbox{6-31G(d)} as some examples of particularly efficient basis sets
 for calculating the geometry in RHF-heterolevel MCs.  Under the
 approximation in sec.~\ref{subsec:pess_distance}, they can be used
 without altering the relevant conformational behaviour of
 polypeptides of more than a hundred residues (i.e., their
 distance~$d_{12}$ with the homolevel \mbox{6-311++G(2df,2pd)} is less
 than $0.1 RT$), and their computational cost is less than 20\% that
 of the reference calculation.
\end{itemize}

Now, after the geometry, we shall investigate the efficiency for
performing energy calculations within RHF of the all the basis sets in
table~\ref{tab:basis_sets} but the largest one. To render the study
meaningful, the geometry on top of which the single-points are
computed must be the same, and we have chosen it to be the one
calculated at the level
\mbox{$L^{\mathrm{best}}:=$6-311++G(2df,2pd)}. Of course, since the
reference to which the $L_{E}^i/\!/L_G^{\mathrm{best}}$ heterolevel
MCs must be compared is the $L^{\mathrm{best}}$ homolevel, and they
take more computational time than this MC (the time
$t_{\mathrm{best}}$ plus the one required to perform the single-point
at $L_{E}^i$), \emph{all of them are computationally inefficient a
priori}. Therefore, in the efficiency plots in
fig.~\ref{fig:distance_hetero_sp_RHF}, the time shown in the $y$-axis
is not the one needed to calculate the actual PES with the
$L_{E}^i/\!/L_G^{\mathrm{best}}$ MC, but just the one
required for the single-point computation. In principle therefore, the
study and the conclusions drawn should be regarded only as providing
\emph{hints} about how efficient a given basis set will be if it is
used to calculate the energy on top of some less demanding geometry
than the $L^{\mathrm{best}}$ one (in order to have a MC
that could have some possibility of being efficient). However, in the
fourth part of the RHF$/\!/$RHF-intramethod investigation (see below), we
show that the performance of the different basis sets for single-point
calculations depends weakly on the underlying geometry, so that the
range of validity of the present part of study must be thought to be
wider.

In fig.~\ref{fig:distance_hetero_sp_RHF}a, a general picture of the
comparison is presented, whereas, in
fig.~\ref{fig:distance_hetero_sp_RHF}b, a detailed zoom of the most
efficient region of the plot is shown. As we have already mentioned,
the time $t$ shown is the average one per point required to perform
the single-point energy calculation on the best geometry, and,
consequently, the time $t_{\mathrm{best}}$ used for defining the
efficient region has been redefined as the one needed for a
single-point at~$L^{\mathrm{best}}$ (i.e., $t_{\mathrm{best}}\simeq
1.7$ hours).

We extract the following conclusions from the plots:

\begin{itemize}
\item Regarding the check of rules (ii) and (iii), \emph{the situation
 is the same as in the two former cases}, with the only difference
 that we can see that, for single-point calculations,
 6-31G(f,d) is much more inefficient than for geometry optimizations,
 being of an accuracy close to that of the smallest basis set studied,
 the 3-21G, and taking considerably more time.
\item \emph{The} 3-21G \emph{family of basis sets is very inefficient
 for energy calculations}.
\item On the other hand, like it happened in the homolevels case,
 \emph{the} 4-31G \emph{basis sets performance is quite close to that
 of the} \mbox{6-31G} \emph{family}. This suggests that, for energy
 calculations in RHF$/\!/$RHF MCs, to use a considerable
 number of primitive Gaussian shells to form the contracted ones is
 more important in the valence orbitals than in the core ones.
\item \emph{The 1st-row atoms polarization gap in the
 distance} $d_{12}$ \emph{also occurs for single-point calculations}
 (see fig.~\ref{fig:distance_hetero_sp_RHF}a). This time,
 3-21G(d,p) is placed above~it.
\item \emph{The relative inefficiency of the the basis sets with extra
 polarizations}, (2d,2p) \emph{or} (2df,2pd), \emph{and no diffuse
 functions is also observed here for energy calculations}. It is mild,
 like in the homolevels case, and no gap appears.
\item Like in the two studies above, \emph{the addition of diffuse
 functions to singly-polarized} ((d) or (d,p)) \emph{basis sets
 increases the accuracy} for single-point energy calculations as well.
\item About the accuracy of the investigated MCs, the situation
 observed in the homolevel case is even more severe here, since not
 even the 6-31++G(2df,2pd) single-point MC lies in the protein region
 and the worst basis sets (see 3-21G, for example) present distances
 over $3 RT$. This enriches and supports the ideas that underlie the
 \emph{heterolevel assumption}, showing that, \emph{whereas the level
 of the theory may be lowered in the calculation of the (constrained)
 equilibrium geometries, it is necessary to perform high-level energy
 single-points if a good accuracy is sought}.
\item Related to the basis set convergence issue, the situation here is
 analogous to the one seen in the case of homolevel MCs: \emph{No
 accumulation point is reached}, and the accuracy can always be
 increased by intelligently enlarging the basis set.
\item Finally, let us mention 6-311+G(d,p), 6-31+G(d) and
 \mbox{6-31G(d)} as some examples of particularly efficient basis sets
 also for calculating the energy in RHF-heterolevel MCs.
 Under the approximation in sec.~\ref{subsec:pess_distance}, they can
 be used without altering the relevant conformational behaviour of
 short peptides, and their computational cost is less than 10\% that
 of the reference single-point calculation.
\end{itemize}

Next, in order to close the RHF$/\!/$RHF-intramethod section, we
evaluate a group of heterolevel MCs which are
constructed by simultaneously decreasing the level of the theory used
for the geometry and the one used for the energy single-point,
relatively to the reference 6-311++G(2df,2pd).

\begin{figure}
\begin{center}
\includegraphics[scale=0.38]{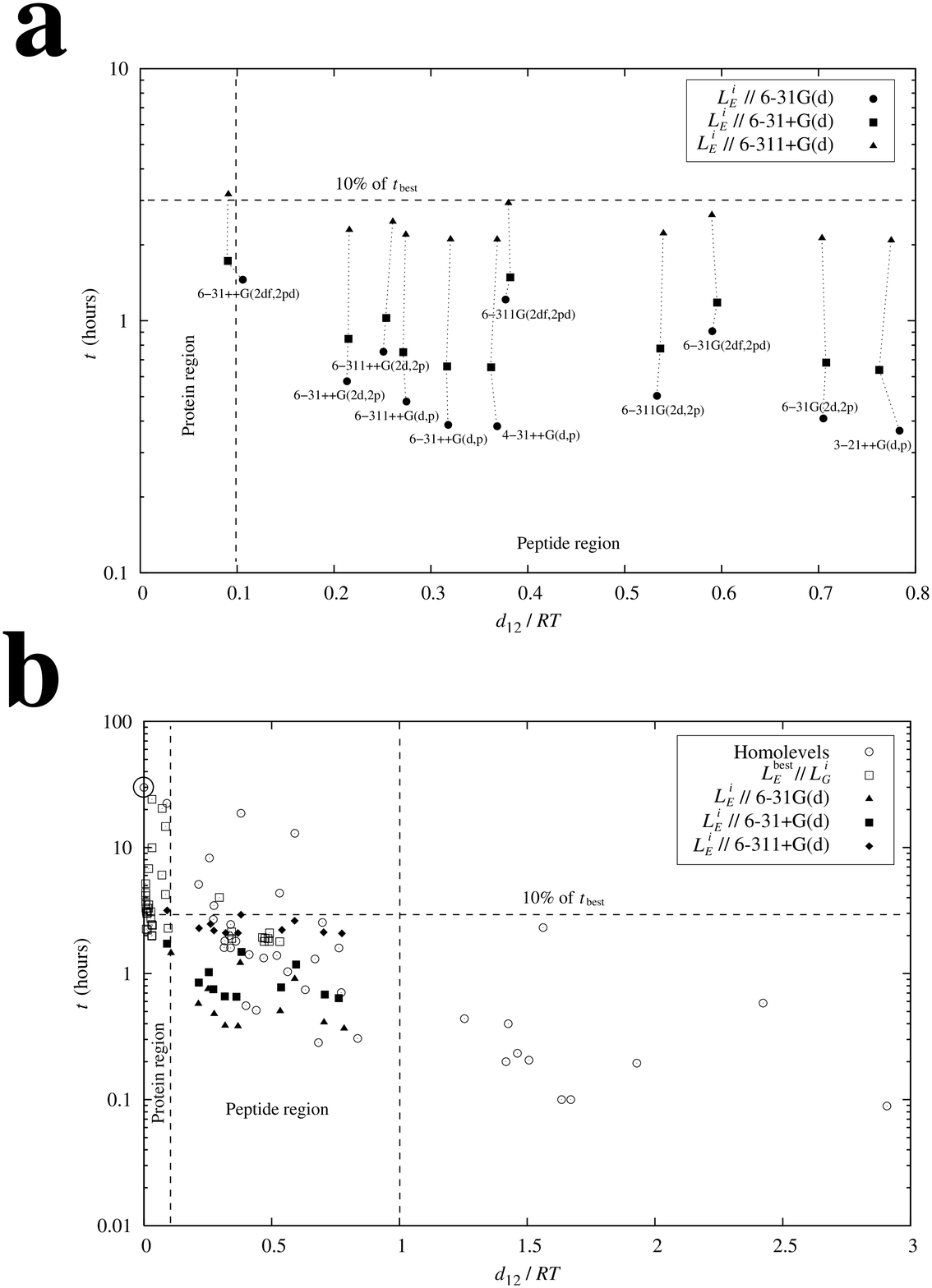}
\caption{\label{fig:distance_selected_and_everything_RHF} {\bf (a)}
Efficiency plot of some selected \emph{RHF-heterolevel} MCs $L_E^i
/\!/ L_G^i$ with $L_E^i \neq L_G^i$ and both of them different from
the best level 6-311++G(2df,2pd). The MCs calculated on top of the
same geometry are joined by broken lines. {\bf (b)} Efficiency plot of
all the MCs in figs.~\ref{fig:distance_homo_RHF},
\ref{fig:distance_hetero_geometry_RHF} and in the (a)-part of this
figure. In both figures, in the $x$-axis, we show the distance
$d_{12}$, in units of $RT$ at \mbox{$300^{\mathrm{o}}$ K}, between any
given MC and the reference one (the \emph{homolevel}
6-311++G(2df,2pd), indicated by an encircled point in (b)), while, in
the $y$-axis, we present in logarithmic scale the average
computational time per point of the 12$\times$12 grid defined in the
Ramachandran space of the model dipeptide HCO-{\small
L}-Ala-NH$_2$. The different accuracy regions depending on $d_{12}$,
are labeled, and the 10\% of the time $t_{\mathrm{best}}$ taken by the
reference homolevel 6-311++G(2df,2pd) is also indicated.}
\end{center}
\end{figure}

Using the basis sets in table~\ref{tab:basis_sets}, there exist $38
\times (38 - 1) = 1406$ different MCs of the form
$L_E^i/\!/L_G^i$, with $L_E^i \neq L_G^i$ and excluding
6-311++G(2df,2pd). This number is too large to perform an exhaustive
study and, therefore, any investigation of the MCs in this particular
group must be necessarily exploratory. Here, we are specially
interested in the most efficient MCs, so that, using the lessons
learned in the preceding paragraphs, we have considered only
heterolevels with $L_G^i$ being 6-31G(d), 6-31+G(d) or 6-311+G(d),
which we have proved to perform well at least when the single-point is
calculated at \mbox{6-311++G(2df,2pd)}. For the choice of $L_E^{i}$,
different criteria have been followed. On the one hand, since the
energy at level $L_G^i$ is readily available as an output of the
geometry optimization step, it is clear that to perform a single-point
calculation with a level of similar accuracy to~$L_G^i$ will not
pay. On the other hand, some hints may be extracted from the study in
fig.~\ref{fig:distance_hetero_sp_RHF} about which could be the most
efficient basis sets for calculating the energy. Taking these two
points into consideration, and also including, for checking purposes,
some levels that are expected to be inefficient, the basis sets that
are investigated for performing single-points within RHF are those
shown in fig.~\ref{fig:distance_selected_and_everything_RHF}a, where
$t_{\mathrm{best}}$ is again the time taken by the reference homolevel
6-311++G(2df,2pd).

\begin{table}[!t]
\begin{center}
\begin{tabular}{l@{\hspace{20pt}}crr}
 Efficient RHF$/\!/$RHF MCs & $d_{12}/RT$ $^{a}$ &
 \multicolumn{1}{c}{$N_{\mathrm{res}}$ $^{b}$} &
 $t$ $^{c}$ \\
\hline\\[-8pt]
 6-311++G(2df,2pd)$/\!/$6-31++G(d,p) & 0.008 & 17382.5 & 11.74\%\\
 6-311++G(2df,2pd)$/\!/$6-31+G(d)    & 0.009 & 11752.4 &  7.53\%\\
 6-311++G(2df,2pd)$/\!/$4-31+G(d)    & 0.014 &  5163.4 &  7.38\%\\
 6-311++G(2df,2pd)$/\!/$6-31G(d)     & 0.031 &  1066.0 &  6.62\%\\[6pt]
 6-31++G(2df,2pd)$/\!/$6-31G(d)      & 0.106 &    89.3 &  4.86\%\\
 6-31++G(2d,2p)$/\!/$6-31G(d)        & 0.213 &    22.0 &  1.92\%\\
 6-311++G(d,p)$/\!/$6-31G(d)         & 0.275 &    13.2 &  1.60\%\\
 6-31++G(d,p)$/\!/$6-31G(d)          & 0.318 &     9.9 &  1.29\%\\
 4-31++G(d,p)$/\!/$6-31G(d)          & 0.368 &     7.4 &  1.27\%\\[6pt]
 6-31G(d)$/\!/$6-31G(d)              & 0.683 &     2.1 &  0.95\%\\
 6-31G$/\!/$6-31G                    & 1.634 &     0.4 &  0.33\%\\
 3-21G$/\!/$3-21G                    & 2.908 &     0.1 &  0.30\%\\
\end{tabular}
\end{center}
\caption{\label{tab:efficient_bs_RHF}{\small List of the most
efficient RHF$/\!/$RHF-intramethod MCs located at the lower-left
envelope of the cloud of points in
fig.~\ref{fig:distance_selected_and_everything_RHF}b. The first block
contains MCs of the form $L_E^{\mathrm{best}}/\!/L_G^i$ (see
fig.~\ref{fig:distance_hetero_geometry_RHF}), the second one those of
the form $L_E^i/\!/L_G^i$ (see
fig.~\ref{fig:distance_selected_and_everything_RHF}a), and the third
one the homolevels in fig.~\ref{fig:distance_homo_RHF}.
$^{a}$Distance with the reference MC (the homolevel
6-311++G(2df,2pd)), in units of $RT$ at \mbox{$300^{\mathrm{o}}$
K}. $^{b}$Maximum number of residues in a polypeptide potential up to
which the corresponding MC may correctly approximate the reference
(under the assumptions in
sec.~\ref{subsec:pess_distance}). $^{c}$Required computer time,
expressed as a fraction of $t_{\mathrm{best}}$.}}
\end{table}

There are no essentially new conclusions to extract from this part of
the study, since it mainly confirms those drawn from the previous
parts and shows that they can be combined rather independently. For
example, the approximate verticality of the dotted lines joining the
MCs with equal $L_E^i$ indicates, as we have already mentioned, that,
in the RHF$/\!/$RHF case, \emph{the accuracy of a given MC depends
much more strongly on the level used for calculating the energy than
on the one used for the geometry}. Also, the fact that the MCs with
$L_G^i=$ 6-31G(d) lie in the lower-left envelope of the plot shows that
6-31G(d) keeps its character of efficient basis set for computing the
geometry even if the single-point is calculated with levels that are
different from the reference one. Finally, note that, in this
particular problem, within RHF, and under the assumptions in
sec.~\ref{subsec:pess_distance}, if one wants to correctly approximate
the reference MC beyond 100-residue peptides, the energy must be
calculated at 6-31++G(2df,2pd).

In fig.~\ref{fig:distance_selected_and_everything_RHF}b, all the 110
MCs studied up to now are depicted as a summary (the 38 inefficient
$L_E^i/\!/L_G^{\mathrm{best}}$ ones are not shown). Now, if we look at
the lower-left envelope of the plot, we can see that, depending on the
target accuracy sought, the most efficient MCs may belong to different
groups among the ones investigated above. From $\sim 0 RT$ to $\sim
0.1 RT$, for example, the most efficient MCs are the
$L_E^{\mathrm{best}}/\!/L_G^i$ ones; from $\sim 0.1 RT$ to $\sim 0.5
RT$, on the other hand, the MCs of the form $L_E^i/\!/L_G^i$, where
the single-point level has also been lowered with respect to the
reference one, clearly outperform those in the rest of groups;
finally, for distances $d_{12} > 0.5 RT$, it is recommendable to use
homolevel MCs. In table~\ref{tab:efficient_bs_RHF}, these efficient
MCs are shown together with their distance $d_{12}$ to the reference
homolevel 6-311++G(2df,2pd), the number of residues $N_{\mathrm{res}}$
up to which they can be used as a good approximation of it, and the
required computer time $t$, expressed as a fraction of
$t_{\mathrm{best}}$.

\subsection{MP2$/\!/$MP2-intramethod model chemistries}
\label{subsec:pess_intraMP2}

\begin{table}[!b]
\begin{center}
\begin{tabular}{llll}
\hline\\[-8pt]
 3-21G   & 6-31G(d)      & 6-31+G(d)      & 6-311+G(d)              \\
 6-31G   & 6-31G(d,p)    & 6-31++G(d,p)   & {\bf 6-311++G(2df,2pd)} \\
 6-31++G & 6-31G(2d,2p)  & 6-31++G(2d,2p) &                         \\[4pt]
\hline
\end{tabular}
\end{center}
\caption{\label{tab:basis_sets_MP2}{\small Basis sets investigated in
the \emph{MP2$/\!/$MP2-intramethod} part of the study. The largest one is
indicated in bold face.}}
\end{table}

Now, using all the information gathered in the previous
RHF$/\!/$RHF-intramethod section (see however
sec.~\ref{subsec:pess_interlude} and the first paragraph of
sec.~\ref{subsec:pess_intraRHF}), we open the second part of the
study, in which we shall perform an \emph{MP2$/\!/$MP2-intramethod}
investigation with some selected basis sets among those in
table~\ref{tab:basis_sets}. The choice of MP2 \cite{Mol1934PR} as the
method immediately `above' RHF is justified by several reasons. In the
first place, it is typically regarded as accurate and as the
reasonable starting point to include correlation in the literature
\cite{DiS2005JCTC,Jen2005ARCC,Hob2002JACS,Hob1999CR,Hal1999JCP,StA1995JCC},
where it is also commonly used as a reference calculation to evaluate
or parameterize less demanding methods
\cite{Zha2005JPCA,San2005CPL,Wan2005CPL,Bor2003JPCB,Bea1997JACS}. Secondly,
and contrarily to DFT, MP2 is a wavefunction-based method that allows
to more or less systematically improve the calculations by going to
higher orders of the M{\o}ller-Plesset perturbation expansion. The
majority of the rest of methods devised to add correlation to the RHF
wavefunction-based results, such as coupled cluster, configuration
interaction, or MCSCF, are more computationally demanding than MP2
\cite{Kno2000PRO,Kut1999AQC,Jen1998BOOK}. Finally, although, for some
particular problems, DFT may rival MP2
\cite{Zha2007JCTC,Tum1999PCCP,Gon1998JCC,StA1995JCC}, the latter is known to
account better for weak dispersion forces, which are present and may
be important in peptides \cite{Kam2007JCTC,vMo2006JPCA,Hob2002JACS}.

\begin{figure}
\begin{center}
\includegraphics[scale=0.20]{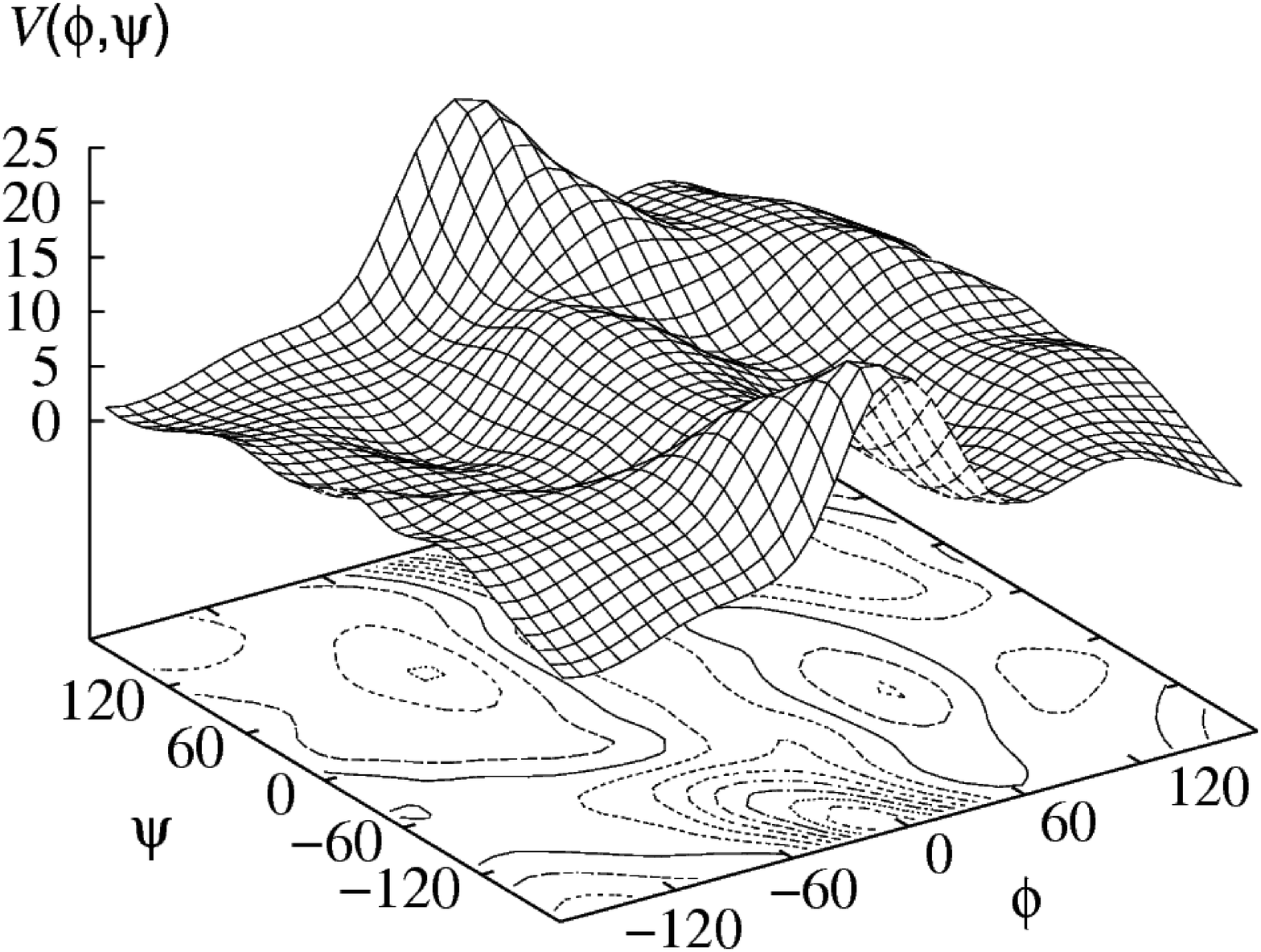}
\caption{\label{fig:best_pes_mp2} Potential energy surface of the
model dipeptide HCO-{\small L}-Ala-NH$_2$ computed at the
MP2/6-311++G(2df,2pd) level of the theory. The PES has been originally
calculated in a 12$\times$12 discrete grid in the space spanned by the
Ramachandran angles $\phi$ and $\psi$ and later smoothed with bicubic
splines for visual convenience. The energy reference has been set to
zero. (At this level of the theory, the absolute energy of the minimum
point in the 12$\times$12 grid, located at $(-75^o,75^o)$, is
$-416.4705201527$ hartree).}
\end{center}
\end{figure}

The basis sets investigated in this MP2 part are the 11 ones in
table~\ref{tab:basis_sets_MP2} and they have been originally chosen in
order to adequately sample the larger set studied at RHF and check if
the same effects are observed at MP2. Some kind of selection must be
done due to the higher computational cost of MP2 calculations, so
that, with the hope that the RHF results were relatively transferable
to MP2, the basis sets that have proved to be relatively more
efficient at RHF were included in table~\ref{tab:basis_sets_MP2},
together with the largest one, the smallest one and a small number of
other basis sets (such as \mbox{6-31G(d,p)} or 6-31G(2d,2p), for
example) intended to analyze the tendencies observed in the previous
section. In the following discussion and in
sec.~\ref{subsec:pess_interlude}, however, the \mbox{RHF $\rightarrow$
MP2} transferability of the results is shown to be imperfect, so that,
despite the valuable lessons learned in this work, in further studies,
one of the research directions that will have to be followed is the
addition of more basis sets to table~\ref{tab:basis_sets_MP2}.

We would also like to stress that the MP2-reference PES of HCO-{\small
L}-Ala-NH$_2$, using 6-311++G(2df,2pd), that has been calculated to
carry out the investigation presented here (see
fig.~\ref{fig:best_pes_mp2}) is, as far as we are aware, \emph{the one
computed at the highest level of the theory at present}.  Although
coupled cluster \cite{Kam2007JCTC,Per2003JCC} and MP4
\cite{Var2002JPCA} methods have been used to perform single-points on
top of the geometries optimized at lower levels for some selected
conformers\footnote{ In this brief review of the literature, we
include, apart from the calculations in HCO-{\small L}-Ala-NH$_2$,
also those in alanine dipeptides with different protecting groups,
since both efficiency and accuracy considerations are expected to be
very similar.}, the highest homolevels used to calculate full PESs in
the literature after the one used in this study seem to be
MP2/6-311G(d,p) in ref.~\citen{Yu2001JMS} and
B3LYP/\mbox{6-311++G(d,p)} in ref.~\citen{Per2003JCC} (assuming that
the accuracy of the B3LYP method lies somewhere between RHF and
MP2). Regarding heterolevel MCs, in ref.~\citen{Mac2004JCC}, a PES is
calculated at the \mbox{LMP2/cc-pVQZ(-g)$/\!/$MP2/6-31G(d)} level, and
although this is certainly a remarkably accurate computation, the
question whether it is better than the homolevel MP2/6-311++G(2df,2pd)
will have to wait until an assessment of the LMP2 method and its
relation to the heterolevel hypothesis is performed. Finally,
something analogous may be said about the
MP2/cc-pVTZ$/\!/$MP2/6-31G(d) PES in ref.~\citen{Wan2004JCC}.

\begin{figure}
\begin{center}
\includegraphics[scale=0.32]{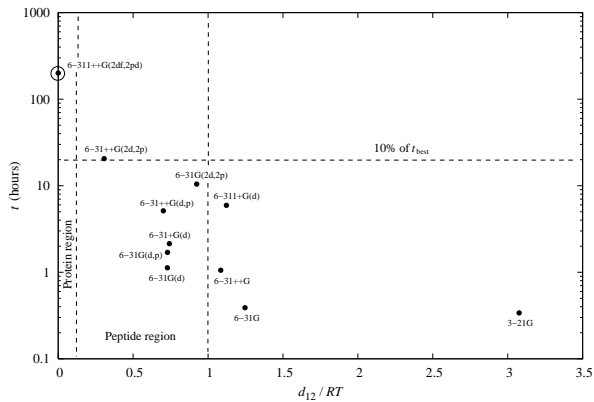}
\caption{\label{fig:distance_homo_MP2} Efficiency plot of the
\emph{MP2-homolevel} MCs corresponding to all the basis
sets in table~\ref{tab:basis_sets_MP2}. In the $x$-axis, we show the
distance $d_{12}$ in units of $RT$, at \mbox{$300^{\mathrm{o}}$ K},
between any given MC and the reference one (the
\emph{homolevel} 6-311++G(2df,2pd), indicated by an encircled point),
while, in the $y$-axis, we present in logarithmic scale the average
computational time taken for each MC, per point of the
12$\times$12 grid defined in the Ramachandran space of the model
dipeptide HCO-{\small L}-Ala-NH$_2$.}
\end{center}
\end{figure}

Now, the structure of the MP2$/\!/$MP2-intramethod study is the same
as in the RHF$/\!/$RHF case: We begin by evaluating the \emph{MP2
homolevels}, and, just as we did before, the `MP2' keyword is omitted
from the MCs specification, since, in this section, no possible
ambiguity may appear.

In fig.~\ref{fig:distance_homo_MP2}, the \emph{homolevel} MCs
corresponding to all the basis sets in table~\ref{tab:basis_sets_MP2}
are compared to the reference one. It takes an average of~$\sim 200$
hours $\simeq$ 8 days of computer time per grid point (see
page~\pageref{foot:howtimegeo}) to calculate the PES of the model
dipeptide HCO-{\small L}-Ala-NH$_2$ at the reference homolevel
6-311++G(2df,2pd); this time is denoted by~$t_{\mathrm{best}}$.

Regarding the conclusions that can be extracted from this plot, let us
focus (remarking the differences) on the issues parallel to the ones
studied in the RHF case, although, since the number of basis sets in
table~\ref{tab:basis_sets_MP2} is smaller than that in
table~\ref{tab:basis_sets}, some details will have to be left out:

\begin{itemize}
\item \emph{The} 3-21G \emph{basis set is again the worst one for
 homolevel calculations}, with a distance close to 3 $RT$.
\item The 1st-row atoms polarization gap that we saw in
 fig.~\ref{fig:distance_homo_RHF}a, \emph{is absent here}, and, for
 example, the 6-31++G basis set is more accurate than the larger and
 polarized \mbox{6-311+G(d)}.
\item \emph{The only basis set with extra polarizations and no diffuse
 functions that we have studied in the MP2 case,} 6-31G(2d,2p)\emph{,
 is less efficient than its diffuse functions-containing counterpart,}
 6-31++G(2d,2p).
\item Whereas, in the RHF case, the addition of diffuse functions to
 singly-polarized ((d) or (d,p)) basis sets always increased the
 accuracy, here, it is sometimes slightly advantageous (in the
 6-31G(d,p) $\rightarrow$ 6-31++G(d,p) case) and sometimes slightly
 disadvantageous (in the 6-31G(d) $\rightarrow$ 6-31+G(d) case). So that
 no clear conclusion may be drawn to this respect.
\item \emph{There is no basis set whose homolevel MC lies in the
 protein region}, although we remark that the second largest basis set
 studied with RHF, 6-31++G(2df,2pd), which lied in the protein
 region then, has not been included in this MP2 part of the work.
\item If we look at the most efficient basis sets (those that lie at
 the lower-left envelope of the `cloud' of points), we can see that,
 like in RHF, \emph{no accumulation point is reached}, i.e., that,
 although the distance between 6-311++G(2df,2pd) and
 \mbox{6-31++G(2d,2p)} is small enough to suggest that we are close to
 the MP2 limit for this particular problem, if the basis set is
 intelligently enlarged, we obtain increasingly better MCs. Also note
 that, if we compare fig.~\ref{fig:distance_homo_MP2} here to
 fig.~\ref{fig:distance_homo_RHF} in the previous section, we do not
 observe a strong signal indicating the slower basis set convergence
 of the MP2 method that is sometimes reported in the literature
 \cite{Bac2007JCP,Mul2006TR,Jen1998BOOK}. Therefore,
 from these limited data, we must conclude that, \emph{for
 conformational energy differences in peptides, the homolevel MCs
 converge approximately at the same pace towards the infinite basis
 set limit for RHF and MP2}.
\item For less than 10\% the cost of the reference calculation, some
 particularly efficient basis sets for MP2-homolevel MCs
 that can be used without altering the relevant conformational
 behaviour of short peptides (i.e., whose distance $d_{12}$ with
 6-311++G(2df,2pd) is less than $RT$) are 6-31++G(d,p), 6-31G(d,p) and
 \mbox{6-31G(d)}.
\end{itemize}

\begin{figure}
\begin{center}
\includegraphics[scale=0.27]{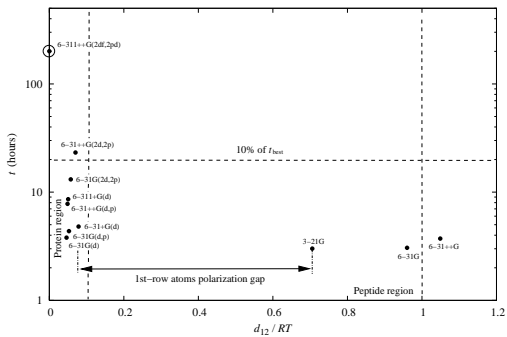}
\caption{\label{fig:distance_hetero_geometry_MP2} Efficiency plot of
the \emph{MP2-heterolevel} MCs $L_E^{\mathrm{best}} /\!/ L_G^{i}$
obtained computing the geometries with all the basis sets in
table~\ref{tab:basis_sets_MP2} but the largest one and then performing
a single-point energy calculation at the best level of the theory,
$L^{\mathrm{best}}:=$6-311++G(2df,2pd), on top of each one of them. In
the $x$-axis, we show the distance $d_{12}$, in units of $RT$ at
\mbox{$300^{\mathrm{o}}$ K}, between any given MC and the reference
one (the \emph{homolevel} 6-311++G(2df,2pd), indicated by an encircled
point), while, in the $y$-axis, we present in logarithmic scale the
average computational time taken for each MC, per point of the
12$\times$12 grid defined in the Ramachandran space of the model
dipeptide HCO-{\small L}-Ala-NH$_2$.}
\end{center}
\end{figure}

Next, in fig.~\ref{fig:distance_hetero_geometry_MP2}, the reference
homolevel 6-311++G(2df,2pd) is compared to the
\emph{MP2$/\!/$MP2-intramethod-heterolevel} MCs $L_E^{\mathrm{best}}
/\!/ L_G^{i}$ obtained computing the geometries with the~10 remaining
basis sets in table~\ref{tab:basis_sets_MP2} and then performing a
single-point energy calculation at the best level of the theory,
$L^{\mathrm{best}}:=$6-311++G(2df,2pd), on top of each one of the
structures. Like in the RHF case, the aim of this comparison is
twofold: on the one hand, we want to measure the relative efficiency
of the different basis sets for calculating the \emph{geometry} (not
the energy), on the other hand, we want to find out whether or not the
\emph{heterolevel assumption} described in the introduction is a good
approximation within MP2.

The average time per point $t$ of the heterolevel MCs has been
calculated adding the average cost of performing a single-point at
$L^{\mathrm{best}}:=$ 6-311++G (2df,2pd) ($\sim 2.7$ hours) to the
average time per point needed to calculate the geometry at each one of
the levels $L_{G}^i$ (see page~\pageref{foot:howtimegeo}).

The following remarks may be made about
fig.~\ref{fig:distance_hetero_geometry_MP2}:

\begin{itemize}
\item Although the only representant of the 3-21G family of basis sets
 in this MP2$/\!/$MP2-intramethod study is one of the most inaccurate
 levels for calculating the geometry, the signal observed in the RHF
 case, indicating that the 3-21G basis sets are not so bad to account
 for the geometry, \emph{also occurs here}, where we can see that
 3-21G is more accurate (and hence more efficient) than the larger
 6-31G and 6-31++G.
\item Contrarily to the homolevel case, here we can appreciate, like
 we did in RHF, a rather wide \emph{gap} in the values of the distance
 $d_{12}$ separating the MCs with the geometry calculated using basis
 sets that contain 1st-row atoms polarization functions
 from those that do not.
\item The signal noticed in the homolevel case regarding the relative
 inefficiency of the the basis sets with extra polarizations and no
 diffuse functions \emph{has been inverted here}, since the
 6-31++G(2d,2p) is less accurate than the smaller 6-31G(2d,2p).
\item Again, and contrarily to the RHF case, the addition of diffuse
 functions to singly-polarized ((d) or (d,p)) basis sets it is
 sometimes slightly advantageous (in the \mbox{6-31G(d,p)}
 $\rightarrow$ 6-31++G(d,p) case) and sometimes slightly
 disadvantageous (in the 6-31G(d) $\rightarrow$ 6-31+G(d) case). So
 that no clear conclusion may be drawn to this respect.
\item Like in the RHF case, and contrarily to the situation for MP2
  homolevels, where no basis sets lied in the protein region and some
  MCs presented distances of near $3 RT$ with the reference one, here,
  most MCs lie well below \mbox{$d_{12} = RT$}, and those for which
  the geometry has been computed with a basis set that contains
  1st-row atoms polarization functions are \emph{all in the protein
  region}, so that, under the assumptions in
  sec.~\ref{subsec:pess_distance}, they can correctly approximate the
  reference~MC for chains of more than 100 residues. Remarkably, some
  of these heterolevel~MCs, such as 6-311++G(2df,2pd)$/\!/$6-31G(d)
  for example, \emph{are physically equivalent to the reference
  homolevel up to peptides of 400 residues at less of~10\% the
  computational cost}. Indeed, all these results \emph{confirm the
  heterolevel assumption}, discussed in the introduction and so
  commonly used in the
  literature\cite{Kam2007JCTC,Cur2007JCP,Arn2006JPCB,San2005CPL,Wan2004JCC,Per2003JCC,Iwa2002JMST,Hob2002JACS,Kam2001JPCB},
  for MP2$/\!/$MP2-intramethod MCs.
\item Differently from the homolevel case, \emph{an accumulation point
  is reached} here in the basis sets, since we can see that there is
  no noticeable increase in accuracy beyond 6-31G(d). Regarding the
  convergence towards the infinite basis set limit, we observe again
  that, whereas it is slightly slower here than in
  fig.~\ref{fig:distance_hetero_geometry_RHF}, the signal is too weak
  to conclude anything and we repeat what we said in the homolevel
  case: that, \emph{for conformational energy differences in peptides,
  the ability of accounting for the geometry in heterolevel MCs of the
  form} $L_E^{\mathrm{best}} /\!/ L_G^{i}$ \emph{converges
  approximately at the same pace towards the infinite basis set limit
  for RHF and MP2}.
\item Finally, let us mention 6-31G(d) as the one clear example of a
 particularly efficient basis set for calculating the geometry in
 MP2-heterolevel MCs. Under the assumptions in
 sec.~\ref{subsec:pess_distance}, it can be used without altering the
 relevant conformational behaviour of polypeptides of around 400
 residues (i.e., its distance~$d_{12}$ with the homolevel
 \mbox{6-311++G(2df,2pd)} is $\sim 0.05 RT$), and its computational
 cost is $\sim 2\%$ that of the reference calculation. The rest of the
 basis sets in fig.~\ref{fig:distance_hetero_geometry_MP2} are either
 less accurate and not significantly cheaper, or more expensive and
 not more accurate than 6-31G(d).
\end{itemize}

\begin{figure}
\begin{center}
\includegraphics[scale=0.28]{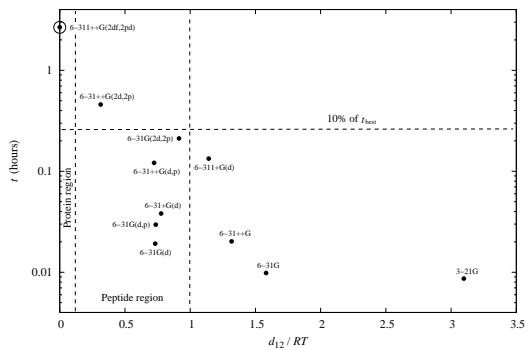}
\caption{\label{fig:distance_hetero_sp_MP2} Efficiency plot of
the \emph{MP2-heterolevel} MCs $L_E^{i} /\!/
L_G^{\mathrm{best}}$ obtained computing the geometry at the best level
of the theory, $L^{\mathrm{best}}:=$6-311++G(2df,2pd), and then
performing a single-point calculation with all the basis sets in
table~\ref{tab:basis_sets} but the largest one. In the $x$-axis, we
show the distance $d_{12}$, in units of $RT$ at
\mbox{$300^{\mathrm{o}}$ K}, between any given MC and the
reference one (the \emph{homolevel} 6-311++G(2df,2pd), indicated by an
encircled point), while, in the $y$-axis, we present in logarithmic
scale the average computational time taken for the corresponding
single-point, per point of the 12$\times$12 grid defined in the
Ramachandran space of the model dipeptide HCO-{\small L}-Ala-NH$_2$.}
\end{center}
\end{figure}

Next, after the geometry, we investigate the efficiency for performing
energy calculations of all the basis sets in
table~\ref{tab:basis_sets_MP2} but the largest one. Like in the RHF
case, the geometry on top of which the single-points are computed must
be the same, and we have chosen it to be the one calculated at the
level \mbox{$L^{\mathrm{best}}:=$6-311++G(2df,2pd)}.  Again, since the
reference to which the $L_{E}^i/\!/L_G^{\mathrm{best}}$ heterolevel
MCs must be compared is the $L^{\mathrm{best}}$ homolevel, and they
take more computational time than this MC (the time
$t_{\mathrm{best}}$ plus the one required to perform the single-point
at $L_{E}^i$), \emph{all of them are computationally inefficient a
priori}. Therefore, in the efficiency plot in
fig.~\ref{fig:distance_hetero_sp_MP2}, the time shown in the $y$-axis
is not the one needed to calculate the actual PES with the
$L_{E}^i/\!/L_G^{\mathrm{best}}$ MC, but just the one required for the
single-point computation. In principle therefore, the study and the
conclusions drawn should be regarded only as providing \emph{hints}
about how efficient a given basis set will be if it is used to
calculate the energy on top of some less demanding geometry than the
$L^{\mathrm{best}}$ one (in order to have a MC that could have some
possibility of being efficient). However, in the fourth part of the
MP2$/\!/$MP2-intramethod investigation (see below), we show, like we
did in the RHF case, that the performance of the different basis sets
for single-point calculations depends weakly on the underlying
geometry, so that the range of validity of the present part of study
must be thought to be wider. Again, the time $t_{\mathrm{best}}$ used
for defining the efficient region in
fig.~\ref{fig:distance_hetero_sp_MP2} has been redefined as the one
needed for a single-point at~$L^{\mathrm{best}}$.

The conclusions of this part of the study are:

\begin{itemize}
\item Like in RHF, 3-21G \emph{is very inefficient
 for energy calculations}.
\item Although all basis sets containing 1st-row atoms
 polarization functions are more accurate than the ones that do not,
 differently from the geometry case, \emph{we do not observe a clear
 gap in the distance} $d_{12}$ \emph{separating the two groups} for
 MP2 single-point energy calculations.
\item Similarly to the MP2-homolevel case and to RHF, the respective
 positions in the plot of \mbox{6-31G(2d,2p)} and 6-31++G(2d,2p)
 constitute a signal that indicates that \emph{the basis sets with
 extra polarizations and no diffuse functions are less efficient than
 their diffuse functions-containing counterparts for energy
 calculations}.
\item As in the rest of the MP2 study, nothing conclusive can be said
 about the addition of diffuse functions to the singly polarized
 6-31G(d) and 6-31G(d,p) basis sets.
\item Regarding the accuracy of the investigated MCs, the situation
 here is analogous to the one found for RHF, and, again this supports
 the ideas that underlie the \emph{heterolevel assumption}, showing
 that, \emph{also at MP2, whereas the level of the theory may be
 lowered in the calculation of the (constrained) equilibrium
 geometries, it is necessary to perform high-level energy
 single-points if a good accuracy is sought}.
\item Related to the basis set convergence issue, the situation here
 is analogous to the one seen in the case of homolevel MCs: \emph{No
 accumulation point is reached}, and the accuracy can always be
 increased by intelligently enlarging the basis set. The convergence
 velocity towards the MP2 limit is again very similar to the one in
 RHF.
\item Finally, let us mention 6-31G(d,p) and \mbox{6-31G(d)} as some
 examples of particularly efficient basis sets for calculating the
 energy in MP2-heterolevel MCs. They can be used without altering the
 relevant conformational behaviour of short peptides, and their
 computational cost is less than 10\% that of the reference
 single-point calculation.
\end{itemize}

\begin{figure}
\begin{center}
\includegraphics[scale=0.27]{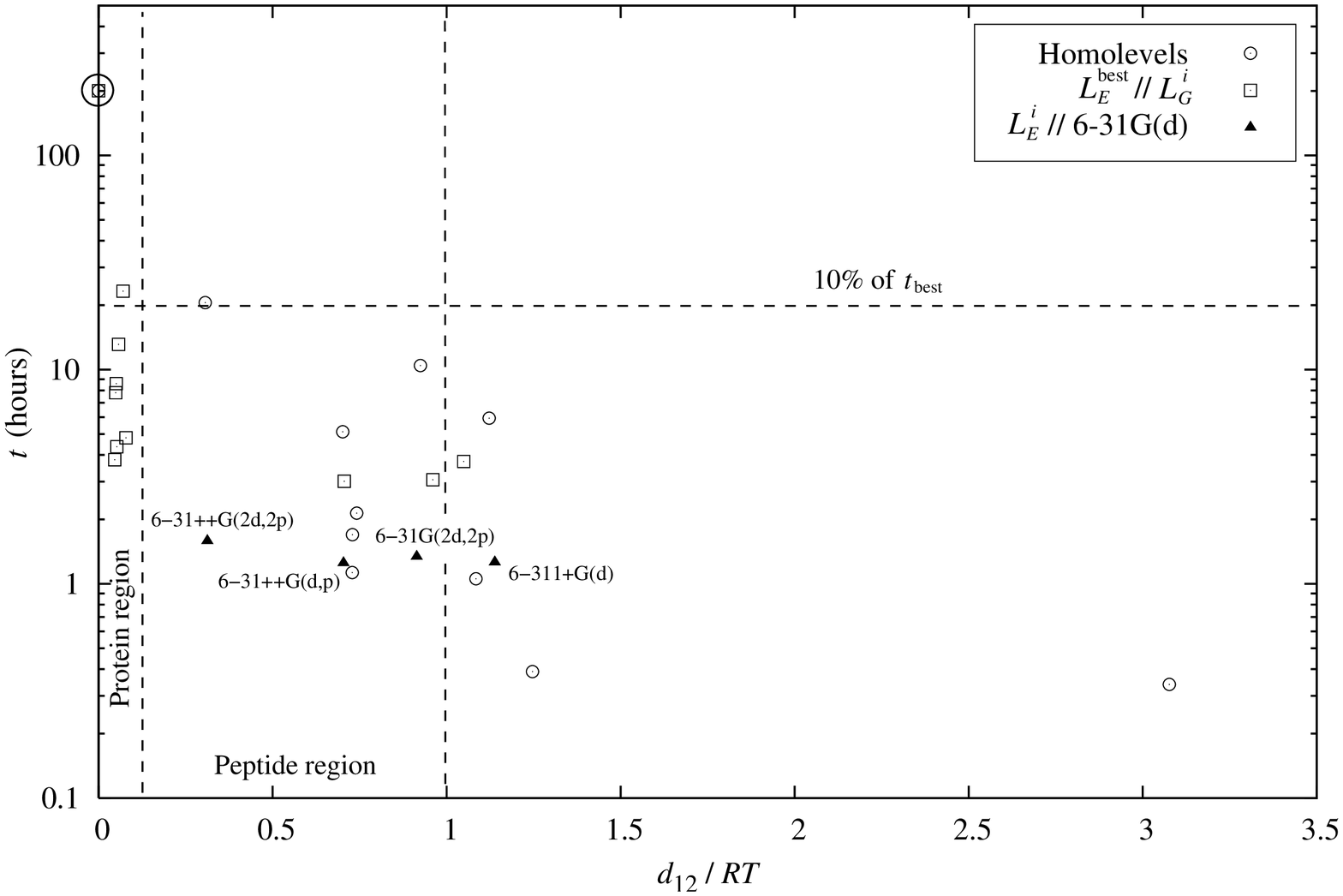}
\caption{\label{fig:distance_selected_and_everything_MP2}
Efficiency plot of all the MCs in
figs.~\ref{fig:distance_homo_MP2} and
\ref{fig:distance_hetero_geometry_MP2}, and also of four additional
ones of the form $L_E^{i}$$/\!/$6-31G(d). Only the latter are labeled. In
the $x$-axis, we show the distance $d_{12}$, in units of $RT$ at
\mbox{$300^{\mathrm{o}}$ K}, between any given MC and the
reference one (the \emph{homolevel} 6-311++G(2df,2pd), indicated by an
encircled point), while, in the $y$-axis, we present in logarithmic
scale the average computational time per point of the 12$\times$12
grid defined in the Ramachandran space of the model dipeptide
HCO-{\small L}-Ala-NH$_2$. The different accuracy regions depending on
$d_{12}$, are labeled, and the 10\% of the time $t_{\mathrm{best}}$
taken by the reference homolevel 6-311++G(2df,2pd) is also
indicated.}
\end{center}
\end{figure}

\begin{table}[!t]
\begin{center}
\begin{tabular}{l@{\hspace{20pt}}crr}
 Efficient MP2$/\!/$MP2 MCs & $d_{12}/RT$ $^{a}$ &
 \multicolumn{1}{c}{$N_{\mathrm{res}}$ $^{b}$} &
 $t$ $^{c}$ \\
\hline\\[-8pt]
 6-311++G(2df,2pd)$/\!/$6-31G(d) & 0.046 &   468.3 &  1.90\%\\[6pt]
 6-31++G(2d,2p)$/\!/$6-31G(d)    & 0.312 &    10.2 &  0.79\%\\
 6-31++G(d,p)$/\!/$6-31G(d)      & 0.703 &     2.0 &  0.62\%\\[6pt]
 6-31G(d)$/\!/$6-31G(d)          & 0.729 &     1.9 &  0.56\%\\
 6-31G$/\!/$6-31G                & 1.247 &     0.6 &  0.19\%\\
 3-21G$/\!/$3-21G                & 3.076 &     0.1 &  0.17\%\\
\end{tabular}
\end{center}
\caption{\label{tab:efficient_bs_MP2}{\small List of the most
efficient MP2$/\!/$MP2-intramethod MCs located at the lower-left
envelope of the cloud of points in
fig.~\ref{fig:distance_selected_and_everything_MP2}. The first block
contains MCs of the form $L_E^{\mathrm{best}}/\!/L_G^i$ (see
fig.~\ref{fig:distance_hetero_geometry_MP2}), the second one those of
the form $L_E^i$$/\!/$6-31G(d) (see
fig.~\ref{fig:distance_selected_and_everything_MP2}), and the third
one the homolevels in fig.~\ref{fig:distance_homo_MP2}.
$^{a}$Distance with the reference MC (the homolevel
\mbox{6-311++G(2df,2pd)}), in units of $RT$ at
\mbox{$300^{\mathrm{o}}$ K}. $^{b}$Maximum number of residues in a
polypeptide potential up to which, under the assumptions in
sec.~\ref{subsec:pess_distance}, the corresponding MC may correctly
approximate the reference. $^{c}$Required computer time, expressed as
a fraction of $t_{\mathrm{best}}$.}}
\end{table}

To close the MP2$/\!/$MP2-intramethod section, we have calculated four
PESs with MCs of the form $L_E^{i}$$/\!/$6-31G(d), since the
geometry computed with 6-31G(d) has proved to be very accurate when
a single-point at the highest level was performed on top of it. Due to
the same computational arguments presented in the previous section,
only those basis sets significantly larger than 6-31G(d) have been
explored for calculating the energy. The results are presented in
fig.~\ref{fig:distance_selected_and_everything_MP2} together with a
summary of the rest of the MP2$/\!/$MP2 MCs studied in this
section (except for the inefficient $L_E^{i} /\!/ L_G^{\mathrm{best}}$
ones).

We have already advanced a conclusion that may be extracted from this
last plot, namely, that if we compare the distance $d_{12}$ of the
$L_E^{i}$$/\!/$6-31G(d) MCs in
fig.~\ref{fig:distance_selected_and_everything_MP2} to the distance of
the $L_E^i/\!/L_G^{\mathrm{best}}$ ones in
fig.~\ref{fig:distance_hetero_sp_MP2} for the same $L_E^i$, we see
that they are very close. Therefore, like in the RHF case, we conclude
that \emph{the accuracy of a given MC depends much more strongly on
the level used for calculating the energy than on the one used for the
geometry}.

Finally, in table~\ref{tab:efficient_bs_MP2}, we present the most
efficient MCs that lie at the lower-left envelope of the plot in
fig.~\ref{fig:distance_selected_and_everything_MP2}. Like in RHF, we
can see that, depending on the target accuracy sought, these most
efficient MCs may belong to different groups among the ones
investigated above. From $\sim 0 RT$ to $\sim 0.1 RT$, for example,
the most efficient MCs are the $L_E^{\mathrm{best}}/\!/L_G^i$ ones;
from $\sim 0.1 RT$ to $\sim 0.75 RT$, on the other hand, the MCs of
the form $L_E^i$$/\!/$6-31G(d) outperform those in the rest of groups;
finally, for distances $d_{12} > 0.75 RT$, it is recommendable to use
homolevel MCs.

\subsection{Interlude}
\label{subsec:pess_interlude}

\begin{figure}[!b]
\begin{center}
\includegraphics[scale=0.18]{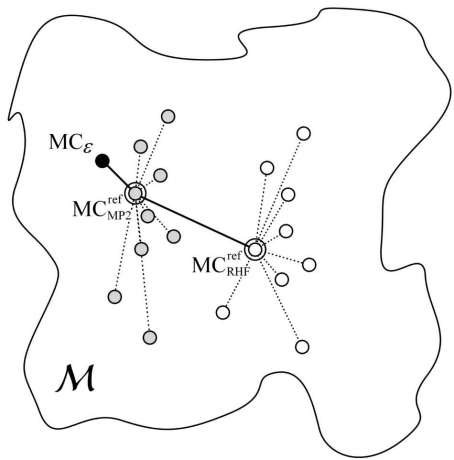}
\caption{\label{fig:mc_space} Space $\mathcal{M}$ of all model
chemistries. The exact model chemistry MC$_\varepsilon$ is shown as a
black circle, MP2 MCs are shown as grey-filled circles and RHF MCs as
white-filled ones. The homolevel reference PESs are indicated with an
additional circle around the points. The situation depicted is
(schematically) the one found in this study.}
\end{center}
\end{figure}

The general abstract framework behind the investigation presented in
this study (and also behind most of the works found in the
literature), may be described as follows:

The objects of study are the \emph{model chemistries} defined by Pople
\cite{Pop1999RMP} and discussed in the introduction. The space
containing all possible MCs is a rather complex and multidimensional
one and it is denoted by $\mathcal{M}$ in fig.~\ref{fig:mc_space}. The
MCs under scrutiny are applied to a particular \emph{problem} of
interest, which may be thought to be formed by three ingredients: the
\emph{physical system}, the \emph{relevant observables} and the
\emph{target accuracy}. The MCs are then selected according to their
ability to yield numerical values of the relevant observables for the
physical system studied within the target accuracy. The concrete
numerical values that one wants to approach are those given by the
\emph{exact model chemistry} MC$_\varepsilon$, which could be thought
to be either the experimental data or the exact solution of the
electronic Schr\"odinger equation. However, the computational effort
needed to perform the calculations required by MC$_\varepsilon$ is
literally infinite, so that, in practice, one is forced to work with a
\emph{reference model chemistry} MC$^\mathrm{ref}$, which, albeit
different from MC$_\varepsilon$, is thought to be close to
it. Finally, the set of MCs that one wants to investigate are compared
to MC$^\mathrm{ref}$ and the nearness to it is seen as approximating
the nearness to MC$_\varepsilon$.

These comparisons are commonly performed using a numerical quantity
$\mathcal{D}$ that is a function of the relevant observables. In order
for the intuitive ideas about relative proximity in the $\mathcal{M}$
space to be captured and the above reasoning to be meaningful, this
numerical quantity $\mathcal{D}$ must have some of the properties of a
mathematical distance. In particular, it is advisable that the
\emph{triangle inequality} is obeyed, so that, for any model chemistry
MC, one has that

\begin{subequations}
\label{eq:pess_ti}
\begin{align}
& \mathcal{D}(\mathrm{MC}_\varepsilon,\mathrm{MC}) \leq
 \mathcal{D}(\mathrm{MC}_\varepsilon,\mathrm{MC}^\mathrm{ref}) +
\mathcal{D}(\mathrm{MC}^\mathrm{ref},\mathrm{MC}) \ ,
  \label{eq:pess_ti_a} \\
\displaybreak[0]
&\mathcal{D}(\mathrm{MC}_\varepsilon,\mathrm{MC}) \geq
 \big| \mathcal{D}(\mathrm{MC}_\varepsilon,\mathrm{MC}^\mathrm{ref}) -
       \mathcal{D}(\mathrm{MC}^\mathrm{ref},\mathrm{MC}) \big| \ ,
  \label{eq:pess_ti_b}
\end{align}
\end{subequations}

and, assuming that
$\mathcal{D}(\mathrm{MC}_\varepsilon,\mathrm{MC}^\mathrm{ref})$ is
small (and $\mathcal{D}$ is a positive function), we obtain

\begin{equation}
\label{eq:pess_ti3}
\mathcal{D}(\mathrm{MC}_\varepsilon,\mathrm{MC}) \simeq
  \mathcal{D}(\mathrm{MC}^\mathrm{ref},\mathrm{MC}) \ ,
\end{equation}

which is the sought result in agreement with the ideas stated at
the beginning of this section.

The distance $d_{12}$ introduced in ref.~\citen{Alo2006JCC} and
summarized in sec.~\ref{subsec:pess_distance}, measured in this case
on the conformational energy surfaces (the relevant observable) of the
model dipeptide HCO-{\small L}-Ala-NH$_2$ (the physical system),
approximately fulfills the triangle inequality and thus captures the
\emph{nearness} concept in the space $\mathcal{M}$ of model
chemistries.

Now, as we have advanced and after having completed the intramethod
parts of the study with both the RHF and MP2 methods, we shall use the
ideas discussed above to tackle the natural question about the
\emph{transferability} of the RHF results to the more demanding and
more accurate MP2-based MCs.

As a first step to answer this question, we point out that the
distance between the reference RHF/6-311++G(2df,2pd) MC and the MP2
one depicted in fig.~\ref{fig:best_pes_mp2} is~$\sim 1.42 RT$. This
prevents us from using the former as an approximation of the latter
even for dipeptides if we want that the conformational behaviour at
room temperature be unaltered. It also indicates that, whereas basis
set convergence has been reasonably achieved within the family of
Pople's Gaussian basis sets, both for homo- and heterolevel MCs and
inside the two methods, the \emph{convergence in method has not been
achieved in the} RHF $\rightarrow$ MP2 \emph{step, even with the
largest basis set investigated} 6-311++G(2df,2pd).

Complementarily to this, in
fig.~\ref{fig:distance_allRHF_correlation_MP2RHF}, we show the
distance of all RHF$/\!/$RHF MCs studied in
sec.~\ref{subsec:pess_intraRHF} (except for the inefficient
$L_E^i/\!/L_G^{\mathrm{best}}$ ones), with both the RHF reference (in
the $y$-axis) and the MP2 one (in the $x$-axis). Some relevant remarks
may be made about the situation encountered:

\begin{figure}
\begin{center}
\includegraphics[scale=0.32]{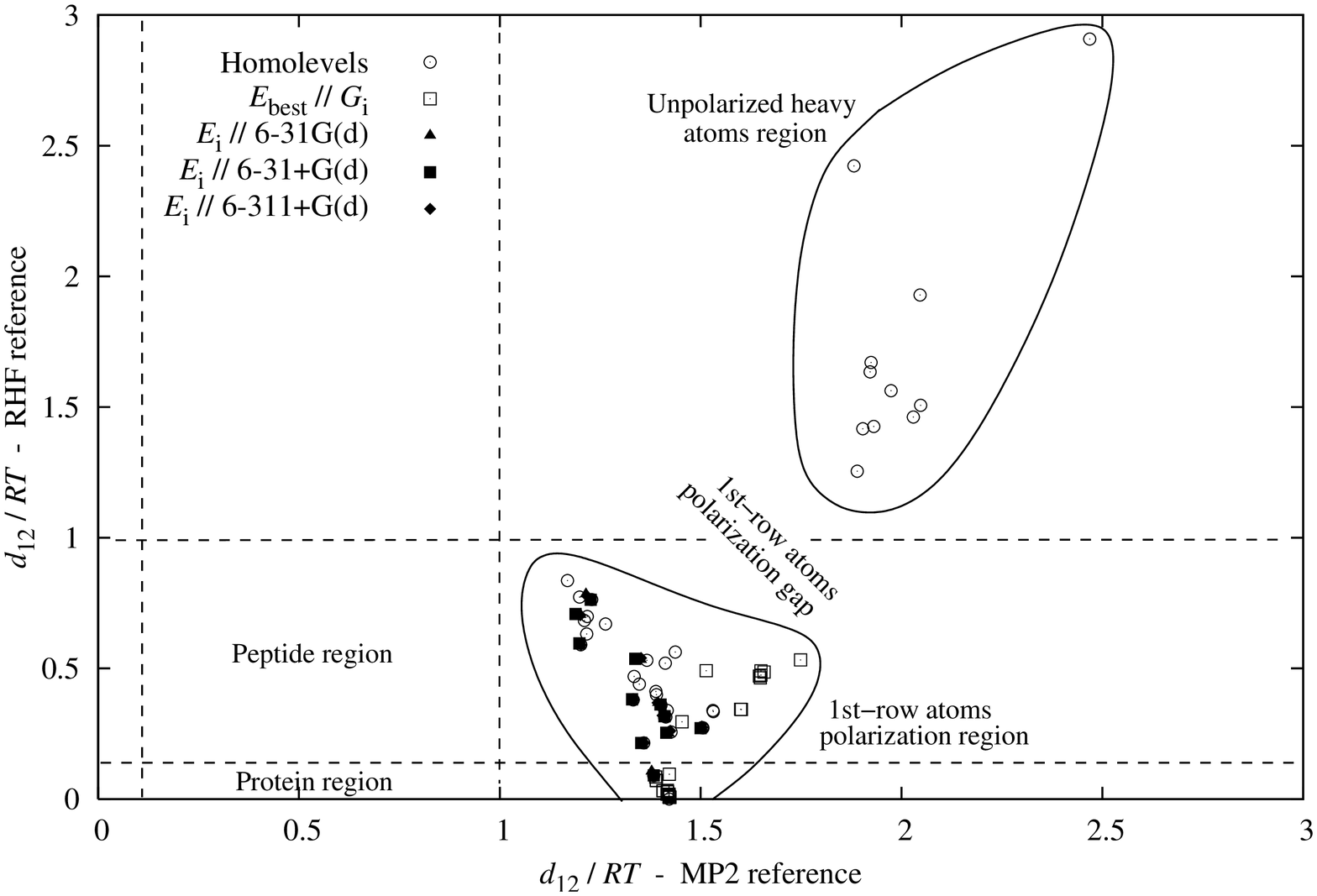}
\caption{\label{fig:distance_allRHF_correlation_MP2RHF} All
\emph{RHF-intramethod} MCs studied in
sec.~\ref{subsec:pess_intraRHF}, except for the inefficient
$L_E^i/\!/L_G^{\mathrm{best}}$ ones. The distance $d_{12}$, in units
of $RT$ at \mbox{$300^{\mathrm{o}}$ K}, with the homolevel
\mbox{MP2/6-311++G(2df,2pd)} reference is shown in the $x$-axis, while
the distance with the RHF reference is shown in the $y$-axis. The
different accuracy regions depending on $d_{12}$, are labeled, and two
groups of \emph{homolevel} MCs are distinguished: those that contain
1st-row atoms polarization shells and those that do not.}
\end{center}
\end{figure}

\begin{itemize}
\item \emph{The distance of all RHF-intramethod MCs to
 the MP2 reference is larger than} $RT$, therefore, none of the former
 may be used to approximate the latter, not even in dipeptides.
\item Although a general trend could be perceived and, for example,
 the RHF homolevels can be clearly divided \emph{in both axes} by the
 1st-row atoms polarization gap found in the previous
 sections, \emph{the correlation between the distance to the MP2
 reference and the distance to the RHF one is as low as $r\simeq
 0.66$}, being $r$ Pearson's correlation coefficient.  Therefore,
 almost all details are lost and the accuracy with respect to
 RHF/6-311++G(2df,2pd) cannot be translated into accuracy with respect
 to the MP2 reference.
\item Related to the previous point, \emph{some strange behaviours are
 present}. For example, not only are there RHF$/\!/$RHF MCs that are
 closer to the MP2 reference than RHF/6-311++G(2df,2pd), but the one
 that is closest is the small \mbox{RHF/4-31G(d,p)}
 homolevel. \emph{This is probably caused by fortituous cancellations
 that shall not allow systematization and that may unpredictably vary
 from one problem to another}. Similar compensations have already been
 observed in the literature \cite{Per2003JCC,Hud2001JCC,End1997JMST}.
\item If we denote by $\mathrm{MC}^\mathrm{ref}_\mathrm{MP2}$ the MP2
 reference MC and, by $\mathrm{MC}^\mathrm{ref}_\mathrm{RHF}$, the RHF
 one, we may use eqs.~(\ref{eq:pess_ti}),

\begin{subequations}
\label{eq:pess_tiref}
\begin{align}
& d_{12}(\mathrm{MC}^\mathrm{ref}_\mathrm{MP2},\mathrm{MC}) \leq
 d_{12}(\mathrm{MC}^\mathrm{ref}_\mathrm{MP2},
   \mathrm{MC}^\mathrm{ref}_\mathrm{RHF}) +
 d_{12}(\mathrm{MC}^\mathrm{ref}_\mathrm{RHF},\mathrm{MC}) \ ,
  \label{eq:pess_tiref_a} \\
\displaybreak[0]
& d_{12}(\mathrm{MC}^\mathrm{ref}_\mathrm{MP2},\mathrm{MC}) \geq
 \big| d_{12}(\mathrm{MC}^\mathrm{ref}_\mathrm{MP2},
         \mathrm{MC}^\mathrm{ref}_\mathrm{RHF}) -
       d_{12}(\mathrm{MC}^\mathrm{ref}_\mathrm{RHF},\mathrm{MC}) \big| \ ,
  \label{eq:pess_tiref_b}
\end{align}
\end{subequations}

 to notice that, since $d_{12}(\mathrm{MC}^\mathrm{ref}_\mathrm{MP2},
 \mathrm{MC}^\mathrm{ref}_\mathrm{RHF})\simeq 1.42 RT$, \emph{for any
 model chemistry MC that is close to the RHF-intramethod reference},
 i.e., that present a small
 $d_{12}(\mathrm{MC}^\mathrm{ref}_\mathrm{RHF},\mathrm{MC})$, we have that

\begin{equation}
\label{eq:pess_tiref3}
d_{12}(\mathrm{MC}^\mathrm{ref}_\mathrm{MP2},\mathrm{MC}) \simeq
  d_{12}(\mathrm{MC}^\mathrm{ref}_\mathrm{MP2},
    \mathrm{MC}^\mathrm{ref}_\mathrm{RHF}) \simeq 1.42 RT \ .
\end{equation}

 This set of RHF-intramethod MCs that are close to the RHF reference,
 and that present the approximately constant value of
 $d_{12}(\mathrm{MC}^\mathrm{ref}_\mathrm{MP2},\mathrm{MC})$ above, can
 be associated to the lower group encircled in
 fig.~\ref{fig:distance_allRHF_correlation_MP2RHF}.
\end{itemize}

All the points above illustrate what we have already advanced at the
beginning of sec.~\ref{subsec:pess_intraMP2}: that the accuracy (or
the efficiency, if computational time is included in the discussion)
of any MC with respect to a good RHF reference, such as the
\mbox{RHF/6-311++G(2df,2pd)} one, \emph{cannot be transferred to
higher levels of the theory} and, therefore, any such comparison must
be seen as providing information only about the infinite basis set
Hartree-Fock limit.

\begin{figure}
\begin{center}
\includegraphics[scale=0.32]{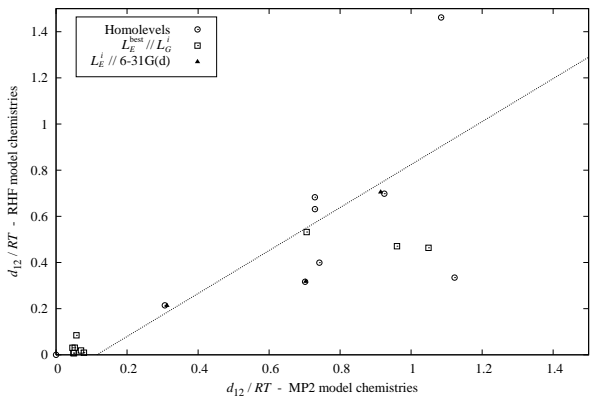}
\caption{\label{fig:distance_allRHF_allMP2_correlation} Distance to
their respective references of the \emph{MP2-} and
\emph{RHF-intramethod} MCs corresponding to the same combination of
basis sets $B_E^i/\!/B_G^i$, expressed in units of $RT$ at
\mbox{$300^{\mathrm{o}}$ K}. Only the region with $d_{12}<1.5RT$ is
shown, and the best-fit line is depicted with a dotted line.}
\end{center}
\end{figure}

To close this section, let us approach the question of the RHF
$\rightarrow$ MP2 transferability of the results from a different
angle.

We have proved in the preceding paragraphs that the study of
RHF-intra\-me\-thod MCs comparing them to a good RHF
reference cannot be used for predicting the accuracy of those MCs with
respect to a probably better MP2 reference. Now, in
sec.~\ref{subsec:pess_intraMP2}, MP2-intramethod MCs
have been compared to the MP2/6-311++G(2df,2pd) homolevel, which, in
turn, has been shown to be close to the infinite basis set MP2
limit. However, this level of the theory is very demanding
computationally: the whole 12$\times$12 grid of points in the PES of
HCO-{\small L}-Ala-NH$_2$ has taken $\sim 3$ years of computer time in
3.20 GHz PIV machines, while the one calculated at
RHF/6-311++G(2df,2pd) has taken `only' $\sim 6$ months (see
sec.~\ref{subsec:pess_QM_calculations}).

Therefore, we have decided to check whether or not the accuracy of a
given RHF-intramethod MC with respect to the RHF reference is
indicative of the accuracy of the MP2-intramethod MC that uses the
same basis sets with respect to its own MP2 reference. The answer to
this question is in
fig.~\ref{fig:distance_allRHF_allMP2_correlation}. There, each point
corresponds to a given combination of basis sets $B_E^i/\!/B_G^i$ and,
in the $x$-axis, the distance between the associated MP2 MC and the
MP2/6-311++G(2df,2pd) reference is shown. In the $y$-axis, on the
other hand, we present the distance of the analogous RHF MC to the
RHF/6-311++G(2df,2pd) homolevel.

Although, since we have had to restrict ourselves to that combinations
that were present both in sec.~\ref{subsec:pess_intraRHF} and in
sec.~\ref{subsec:pess_intraMP2}, the set of MCs is smaller in this
case, the conclusion extracted is that \emph{the correlation is more
significant than before}: $r\simeq 0.92$ if we use all the MCs, and
$r\simeq 0.80$ if we remove the 3-21G homolevel, which is very
inaccurate in both cases, from the set. This indicates that, although
some details might be lost, \emph{the relative efficiency of Gaussian
basis sets in RHF-intramethod studies provides hints about their
performance at MP2}, and it partially justifies the structure of the
investigation presented here.

Finally, the overall situation described in this section and the
relations among all the intramethod MCs studied are schematically
depicted in fig.~\ref{fig:mc_space}.

\subsection{MP2$/\!/$RHF-intermethod model chemistries}
\label{subsec:pess_inter}

\begin{table}[!b]
\begin{center}
\begin{tabular}{llll}
\hline\\[-8pt]
 3-21G   & 6-31G(d)      & 6-31+G(d)      & 6-311+G(d)        \\
 6-31G   & 6-31G(2d,2p)  & 6-31++G(2d,2p) & 6-311++G(2df,2pd) \\[4pt]
\hline
\end{tabular}
\end{center}
\caption{\label{tab:basis_sets_RHFgeo}{\small Basis sets investigated
for calculating the geometry in the \emph{MP2$/\!/$RHF-intermethod} part
of the study.}}
\end{table}

In the final part of the study presented here, we investigate the
efficiency of \emph{heterolevel} MCs in which the
geometry is calculated at RHF and, then, a single-point energy
calculation is performed on top of it at MP2. They shall be termed
\emph{MP2$/\!/$RHF-intermethod MCs}.

To this end, the RHF geometries that are used are those computed with
the 8 basis sets in table~\ref{tab:basis_sets_RHFgeo}. Like in
sec.~\ref{subsec:pess_intraMP2}, they have been selected from those in
table~\ref{tab:basis_sets} looking for the most efficient ones, but
also trying to reasonably sample the whole group of basis sets, in
order to check whether or not the behaviours and signals observed in
the remaining parts of the study are repeated here. The MP2
single-points, on the other hand, are computed with the whole set of
possibilities in table~\ref{tab:basis_sets}.

\begin{figure}
\begin{center}
\includegraphics[scale=0.26]{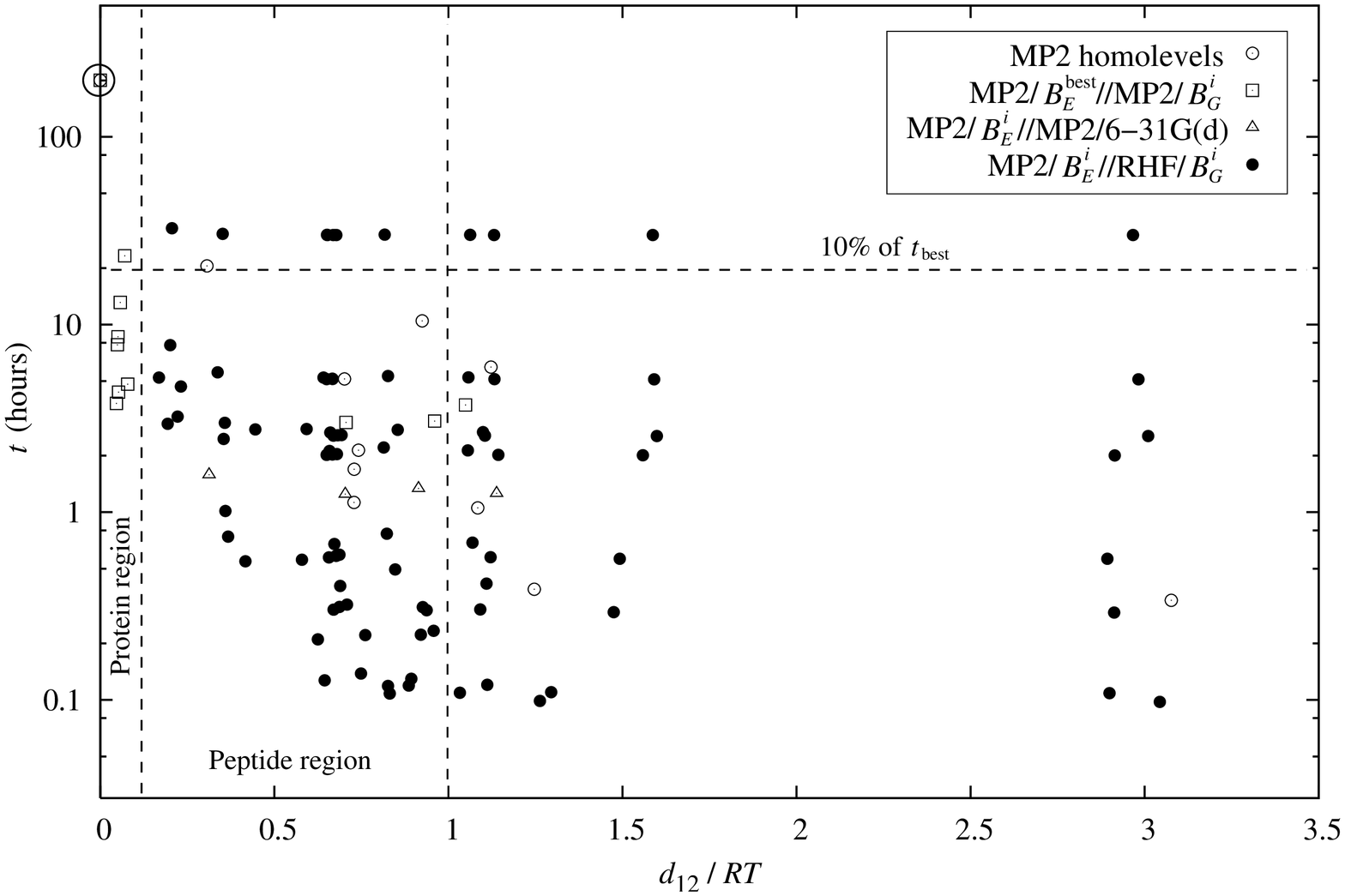}
\caption{\label{fig:distance_allMP2_and_MP2RHF} Efficiency plot of all
the MP2$/\!/$MP2 MCs in
fig.~\ref{fig:distance_selected_and_everything_MP2} together with the
new 88 possible MP2$/\!/$RHF-intermethod combinations of the form
$\mathrm{MP2}/B_E^i/\!/\mathrm{RHF}/B_G^i$ introduced in this
section. In the $x$-axis, we show the distance $d_{12}$, in units of
$RT$ at \mbox{$300^{\mathrm{o}}$ K}, between any given MC and the
reference one (the \emph{homolevel} MP2/6-311++G(2df,2pd), indicated
by an encircled point), while, in the $y$-axis, we present in
logarithmic scale the average computational time per point of the
12$\times$12 grid defined in the Ramachandran space of the model
dipeptide HCO-{\small L}-Ala-NH$_2$. The different accuracy regions
depending on $d_{12}$, are labeled, and the 10\% of the time
$t_{\mathrm{best}}$ taken by the reference homolevel
MP2/6-311++G(2df,2pd) is also indicated.}
\end{center}
\end{figure}

In fig.~\ref{fig:distance_allMP2_and_MP2RHF}, we present an efficiency
plot, using the MP2/6-311++G(2df,2pd) homolevel as reference MC, and
containing all the MP2$/\!/$MP2 MCs studied in
sec.~\ref{subsec:pess_intraMP2} together with the new 88 possible
MP2$/\!/$RHF-intermethod combinations of the form
$\mathrm{MP2}/B_E^i/\!/\mathrm{RHF}/B_G^i$.

Some conclusions can be drawn from this plot:

\begin{itemize}
\item Due to the larger computational demands of the MP2 method, even
 the MCs whose geometry has been computed at the highest RHF level,
 the one with the 6-311++G(2df,2pd) basis set, are much cheaper than
 the MP2 reference. Their times are slightly larger than the 10\% of
 $t_{\mathrm{best}}$, whereas all the rest of MP2$/\!/$RHF MCs take
 less than that bound.
\item For all the RHF geometries, the MCs whose MP2
 single-point has been calculated with 3-21G, 6-31G, 6-31++G and most
 of the 6-311+G(d) ones lie above $d_{12}=RT$, so that they should not
 be used even on dipeptides. This is related to the
 1st-row atoms polarization gap observed in previous
 sections, although the signal is not so strong here.
\item The rest of MP2$/\!/$RHF MCs not included in the two
 previous points lie at the \emph{efficient region}, defined as that
 for which $d_{12}<RT$ and $t<$ 10\% of $t_{\mathrm{best}}$. This
 confirms the \emph{heterolevel assumption} also in the intermethod
 context.
\item However, no MP2$/\!/$RHF-intermethod MCs, not even the ones with
 the single-point calculated at the highest MP2/6-311++G(2df,2pd)
 level, lie in the protein region. Therefore, if we want to
 approximate the reference MP2 results for peptides longer than 100
 residues, under the assumptions in sec.~\ref{subsec:pess_distance},
 \emph{the geometry calculation must be performed at MP2}.  In fact,
 this is the correct way of addressing the discussion between
 Cs{\'a}s{\'a}r \cite{Csa1995JMS} and Sch{\"a}fer
 \cite{Fre1992JACS,Ram1996JMS}, in which the former defends the
 position that the geometry can be calculated at RHF (provided that a
 subsequent MP2 single-point is performed on top of it), while the
 latter disagrees and recommends to compute the geometry at MP2
 too. The data they argue about is, of course, the same; the
 discrepancy simply arises from the fact that Cs{\'a}s{\'a}r is
 thinking only in the small systems in which the calculations have
 been performed, while Sch{\"a}fer wants to use the information
 obtained to gain understanding about the folding of long peptides. In
 this work, the distance introduced in ref.~\citen{Alo2006JCC} and
 summarized in sec.~\ref{subsec:pess_distance} codifies whether or not
 two different MCs yield equivalent PESs for the HCO-{\small
 L}-Ala-NH$_2$ dipeptide (i.e., if $d < RT$, they are physically
 indistinguishable at temperature $T$).  On the other hand, if we
 assume a particular form in which the polypeptide potential is
 constructed from the single-residue PESs and use the additivity
 properties of $d$ \cite{Alo2006JCC}, we can show that this error
 grows with the square root $\sqrt{N}$ of the number of residues and
 use the quantity $N_{\mathrm{res}}$, defined in
 sec.~\ref{subsec:pess_distance} to estimate how the difference
 between the two MCs affects the behaviour of polypeptides. In a work
 in progress in our group \cite{Ech2007UNPb}, we are investigating how
 the distance scales with $N$ for different and more realistic
 hypotheses. It is in this sense, that the statements relying on
 $N_{\mathrm{res}}$ should be regarded as an \emph{estimation}.
\item Finally, let us point out that \emph{there is no accuracy region
 where the MP2-homolevel MCs are more efficient than the rest}.
\end{itemize}

\begin{figure}
\begin{center}
\includegraphics[scale=0.40]{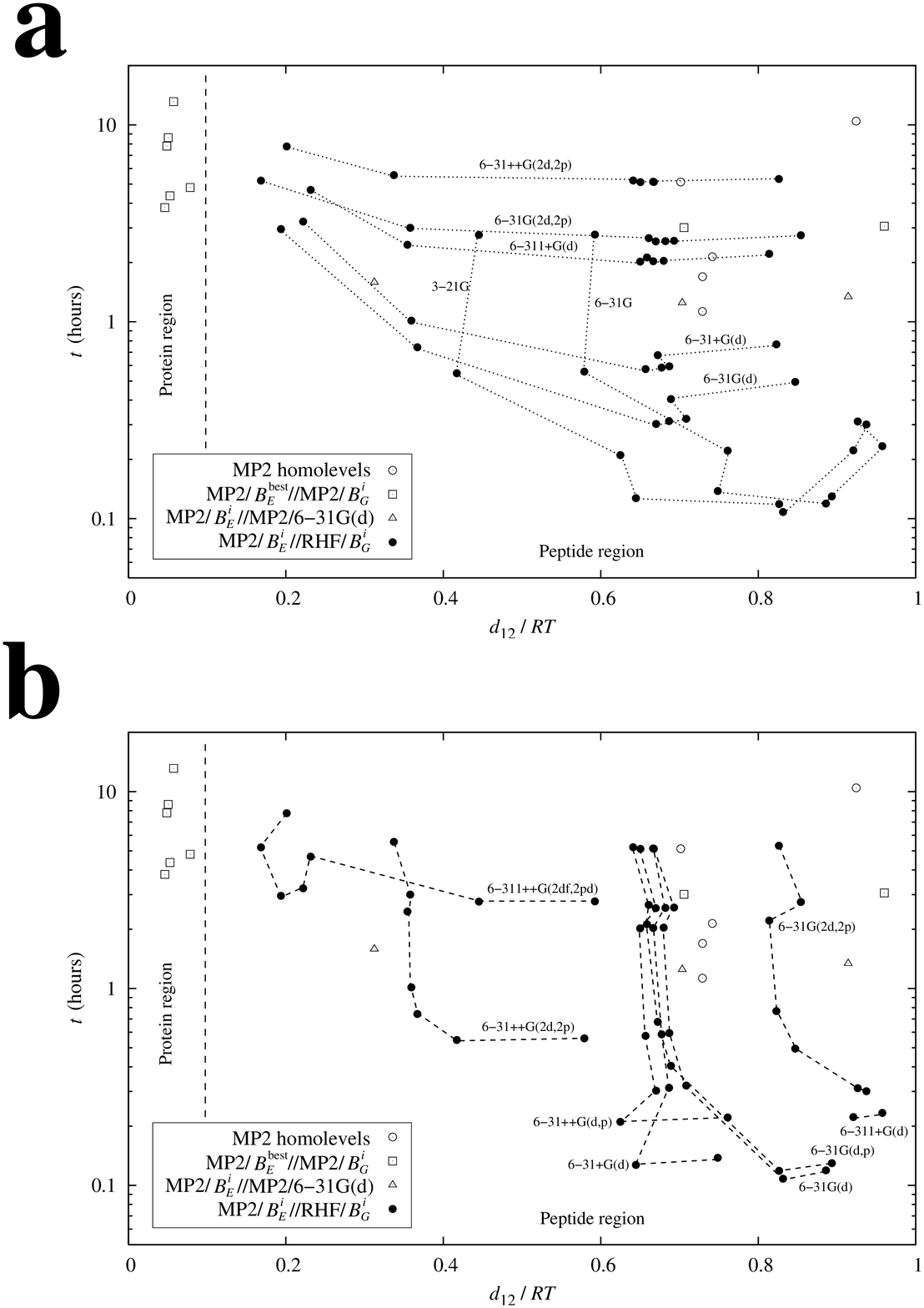}
\caption{\label{fig:distance_allMP2_and_MP2RHF_efficient} Selected
region of the efficiency plot in
fig.~\ref{fig:distance_allMP2_and_MP2RHF}. In {\bf (a)}, the MCs
sharing the same RHF level for the geometry have been joined by
\emph{dotted lines} and the basis set used for that part of the
calculation is indicated.  In {\bf (b)}, the MCs sharing the same MP2
level for the single-point calculations have been joined by
\emph{broken lines} and the corresponding basis set labels are also
shown. The order in which the points have been joined in both cases
has no meaning at all and it is only intended for visual convenience.}
\end{center}
\end{figure}

Now, in fig.~\ref{fig:distance_allMP2_and_MP2RHF_efficient}, the
efficient region of the previous plot is enlarged and, due to the
large number of MP2$/\!/$RHF MCs studied, two subplots are
produced for visual comfort: the one in
fig.~\ref{fig:distance_allMP2_and_MP2RHF_efficient}a, in which the MCs
sharing the same RHF level for the geometry have been joined by dotted
lines, and the one in
fig.~\ref{fig:distance_allMP2_and_MP2RHF_efficient}b, in which the MCs
sharing the same MP2 level for the single-point calculations have been
joined by broken lines.

Let us remark some interesting facts that can be seen in these two
more detailed plots:

\begin{itemize}
\item The leftmost group of five MP2$/\!/$RHF MCs that show the
 highest accuracy are those in which the geometry has been obtained
 with basis sets containing 1st-row atoms polarization functions and
 the single-point energy calculation has been performed at
 MP2/6-311++G(2df,2pd). In particular, the
 MP2/\mbox{6-311++G(2df,2pd)}$/\!/$RHF/6-31G(d) PES can correctly
 approximate the reference one up to peptides of $\sim 25$ residues,
 under the assumptions mentioned above, at around 1\% its
 computational cost. This supports the \emph{heterolevel assumption}
 for MP2$/\!/$RHF-intermethod MCs.

\begin{table}[!t]
\begin{center}
\begin{tabular}{l@{\hspace{20pt}}crr}
 Efficient MP2$/\!/$MP2 and MP2$/\!/$RHF MCs & $d_{12}/RT$ $^{a}$ &
 \multicolumn{1}{c}{$N_{\mathrm{res}}$ $^{b}$} &
 $t$ $^{c}$ \\
\hline\\[-8pt]
 MP2/6-311++G(2df,2pd)$/\!/$MP2/6-31G(d)    & 0.046 &   468.3 &  1.90\%\\[6pt]
 MP2/6-311++G(2df,2pd)$/\!/$RHF/6-31G(d)    & 0.194 &    26.7 &  1.48\%\\
 MP2/6-31++G(2d,2p)$/\!/$RHF/6-31G(d)       & 0.367 &     7.4 &  0.37\%\\
 MP2/6-31++G(2d,2p)$/\!/$RHF/3-21G          & 0.417 &     5.7 &  0.27\%\\
 MP2/6-31+G(d)$/\!/$RHF/3-21G               & 0.645 &     2.4 &  0.06\%\\
 MP2/6-31G(d)$/\!/$RHF/3-21G                & 0.831 &     1.4 &  0.05\%\\[6pt]
 MP2/6-31++G$/\!/$RHF/3-21G                 & 1.033 &     0.9 &  0.05\%\\
 MP2/6-31G$/\!/$RHF/3-21G                   & 1.263 &     0.6 &  0.05\%\\
 MP2/3-21G$/\!/$RHF/3-21G                   & 3.043 &     0.1 &  0.05\%\\
\end{tabular}
\end{center}
\caption{\label{tab:efficient_bs_MP2-RHF}{\small List of the most
efficient MP2$/\!/$MP2 and MP2$/\!/$RHF MCs located at the lower-left
envelope of the cloud of points in
fig.~\ref{fig:distance_allMP2_and_MP2RHF}. The first block contains
the only MP2$/\!/$MP2 MC in the list, the second one the MP2$/\!/$RHF
MCs with a distance $d_{12}$ below $RT$, and the third one those that
are inaccurate even for dipeptides.  $^{a}$Distance with the reference
MC (the homolevel \mbox{MP2/6-311++G(2df,2pd)}), in units of $RT$ at
\mbox{$300^{\mathrm{o}}$ K}. $^{b}$Maximum number of residues in a
polypeptide potential up to which the corresponding MC may correctly
approximate the reference, under the assumptions in
sec.~\ref{subsec:pess_distance}. $^{c}$Required computer time,
expressed as a fraction of $t_{\mathrm{best}}$.}}
\end{table}

\item The RHF geometries calculated with the unpolarized basis sets
 3-21G and 6-31G are, in general, less accurate than the rest, however,
 due to their low computational cost, they turn out to be the most
 efficient ones from $d_{12}\simeq 0.4RT$ on. Remarkably, 3-21G is
 more efficient than 6-31G.
\item In fig.~\ref{fig:distance_allMP2_and_MP2RHF_efficient}b, we can
 observe that, for the medium-sized basis sets 6-31++G (d,p),
 \mbox{6-31+G(d)}, 6-31G(d,p) and 6-31G(d), the single-point accuracy
 is rather insensitive to their differences and they may be used
 interchangeably. There is, however, a weak signal, in the region of
 unpolarized RHF geometries, indicating that the addition of diffuse
 functions may increase the quality of the energy calculations at MP2.
\item The relative accuracy of the MCs whose MP2 single-point has been
 computed at \mbox{6-31++G(2d,2p)} and at 6-31G(2d,2p) suggests that,
 like in previous parts of the study, \emph{it is a good idea to add
 diffuse functions to basis sets that contain doubly-split
 polarizations shells}, also for the MP2 energy calculations in
 MP2$/\!/$RHF-intermethod MCs.
\item Like it happened in sec~\ref{subsec:pess_intraRHF}, in
 fig.~\ref{fig:distance_allMP2_and_MP2RHF_efficient}a, we notice that
 there is no real improvement if we calculate the RHF geometry beyond
 6-31G(d). So that, \emph{an accumulation point is reached for RHF
 geometries in MP2$/\!/$RHF-intermethod MCs}.
\end{itemize}

Finally, in table~\ref{tab:efficient_bs_MP2-RHF}, we present the most
efficient MCs that lie at the lower-left envelope of the plot in
fig.~\ref{fig:distance_allMP2_and_MP2RHF}. These are \emph{the most
efficient MCs found in this work}.

\section{Conclusions}
\label{sec:pess_conclusions}

In this study, we have investigated more than 250 PESs of the model
dipeptide HCO-{\small L}-Ala-NH$_2$ calculated with homo- and
heterolevel RHF$/\!/$RHF, MP2$/\!/$MP2 and MP2$/\!/$RHF MCs. All of
the PESs are available as supplementary material. As far as we are
aware, the highest-level PESs in the literature, the
MP2/6-311++G(2df,2pd) homolevel in fig.~\ref{fig:best_pes_mp2}, has
been used as a reference and all the rest of calculations have been
compared to it (except for sec.~\ref{subsec:pess_intraRHF}, where the
RHF$/\!/$RHF MCs have been compared to RHF/6-311++G(2df,2pd)). The
data and the results extracted are so extense that we have considered
convenient to give here a brief summary of the most important ones.

The first conclusion that we want to point out is that, for the
largest basis set evaluated here, the 6-311++G(2df,2pd) one, for which
the RHF and MP2 limits appear to have been reached, \emph{the
convergence in method has not been achieved}. I.e., the distance
between the MP2 and RHF references is $d_{12}\simeq 1.42RT$, so that
the latter cannot be used to approximate the former even for
dipeptides. Therefore, \emph{we discourage the use of RHF$/\!/$RHF MCs
for peptides}, and, unless otherwise stated, most of the conclusions
below should be understood as referring either to
MP2$/\!/$MP2-intramethod or to MP2$/\!/$RHF-intermethod MCs, which
have proved to be acceptably accurate with respect to the best MP2
calculation.

The second observation related to the comparison between RHF and MP2
is that \emph{the \mbox{RHF $\rightarrow$ MP2} transferability of the
relative accuracies between MCs is imperfect}, and the conclusions
regarding the relative efficiency of the different basis sets arrived
using the former cannot be directly extrapolated to the latter. This
point has two distinct aspects: On the one hand, we have shown that to
compare RHF$/\!/$RHF MCs to a good RHF reference gives little
information about their accuracy with respect to a good MP2 MC. On the
other hand, the comparison with the RHF reference of PESs calculated
with MCs of the form RHF/$B_E^i/\!/$RHF/$B_G^i$ may provide useful
information about the relative accuracy of the analogous
MP2/$B_E^i/\!/$MP2/$B_G^i$ MCs with respect to their own MP2
reference.

Now, keeping these considerations in mind, let us summarize the most
important conclusions pertaining the relative efficiency of the Pople
split-valence basis sets investigated:

\begin{itemize}
\item In the whole study, the polarization shells in 1st-row atoms
 have been shown to be essential to accurately account for both the
 conformational dependence of the geometry and of the energy of the
 system. Except for some particular MCs with 3-21G geometries, which
 may be used if we plan to describe short oligopeptides, \emph{our
 recommendation is that polarization functions in 1st-row atoms be
 included}.
\item In most cases, we have also observed a strong signal indicating
 that \emph{no basis sets should be used containing doubly-split
 polarization shells and no diffuse functions}.
\item \emph{The} 6-31G(d) \emph{basis set}, which is frequently used in the
  literature,
  \cite{Bek2006JACS,vMo2006JPCA,Iwa2002JMST,Per2004JCC,Top2001JACS,Lan2005PSFB,Koo2002JPCA,Bal2000JMS,Els2000CP,Hal1999JCP,Bea1997JACS},
  \emph{has turned out to be a very efficient one for calculating the geometry
    both at RHF and MP2}.
\item Regarding the basis set convergence issue, we can conclude that,
 \emph{for the largest basis sets in the Pople split-valence family,
 both the RHF and MP2 infinite basis set limits are approximately
 reached}.
\item Finally, some \emph{weaker signals} have been observed
 suggesting that to add higher angular momentum polarization shells
 (f,d) before adding the lower ones may be inefficient, that it is not
 recommendable to put polarization or diffuse functions on hydrogens
 only, and that it may be efficient in some cases to add diffuse
 functions to singly-polarized basis sets.
\end{itemize}

Referring to the heterolevel assumption, which, as far as we are
aware, has been tested in this work for the first time in full PESs:

\begin{itemize}
\item As a general and very clear conclusion, since only some
  small-basis set homolevels lie in the lower-left envelope of the
  efficiency plots presented in the previous sections, and, in all
  cases, it happens for distances $d_{12}$ greater than $RT$, we can
  say that \emph{the heterolevel assumption is correct for the
  description of the conformational behaviour of the system studied
  here with MP2$/\!/$MP2 and MP2$/\!/$RHF MCs} (also for
  RHF$/\!/$RHF-heterolevels but, as we remarked above, this has little
  computational interest).
\item Due to the much stronger dependence of the accuracy of MCs on
  the level used for the single-point than on the one used for the
  geometry optimization, together with the lower computational cost of
  the former, \emph{the general recommendation is that the greatest
  computational effort be dedicated to the energy calculation}.
\item Despite this general thumb rule, and under the assumptions in
  sec.~\ref{subsec:pess_distance}, \emph{if one wants to approximate
  the MP2 reference calculation for peptides of more than 100
  residues, the geometry must be calculated using MP2}. Nevertheless,
  with small and cheap basis sets, such as 6-31G(d), the MP2$/\!/$MP2
  results can be good enough at a low computational cost.
\end{itemize}

Finally, let us remark that the investigation performed here has been
done in one of the simplest dipeptides. The fact that we have treated
it as an isolated system, the small size of its side chain and also
its aliphatic character, all play a role in the results
obtained. Hence, for bulkier residues included in polypeptides, and,
specially for those that are charged or may participate in
hydrogen-bonds, the conclusions drawn about the relative importance of
the different type of functions in the basis set, as well as those
regarding the comparison between RHF and MP2, should be approached
with caution and much interesting work remains to be done.

\vspace{0.3cm} All the PESs investigated are publicly available as
supplementary material at {\tt
  http://www.pabloechenique.com/files/public/supp\_materials/}. Each one of
them is a three-column text file containing, in this order, the values of the
Ramachandrand angles $\phi$ and $\psi$ in the 12$\times$12 grid defined in
sec.~\ref{subsec:pess_QM_calculations} and the energy in hartrees. They are
organized in subfolders indicating whether they correspond to RHF- or
MP2-homolevels, or to RHF$/\!/$RHF-, MP2$/\!/$MP2- or
MP2$/\!/$RHF-heterolevels. The filenames are explicative (note that a letter
'o' has been used to indicate that a particular Gaussian shell is missing in
either the 1st-row atoms or the hydrogens).

\section*{Acknowledgments}

\hspace{0.5cm} We would like to thank F. Jensen, T. van Mourik and
A. Perczel for illuminating discussions. The numerical calculations
have been performed at the BIFI supercomputing facilities. We thank
all the staff there, for the invaluable CPU time and the efficiency at
solving the problems encountered.

This work has been supported by the research projects E24/3 and PM048
(Arag\'on Government), MEC (Spain) \mbox{FIS2006-12781-C02-01} and MCyT
(Spain) \mbox{FIS2004-05073-C04-01}. P. Echenique and is supported by
a BIFI research contract.


\end{document}